\documentclass[12pt]{article} 
\usepackage[sectionbib]{natbib}
\usepackage{array,epsfig,fancyheadings,rotating}
\usepackage[]{hyperref}  
\usepackage{sectsty, secdot}
\sectionfont{\fontsize{12}{14pt plus.8pt minus .6pt}\selectfont}
\renewcommand{\theequation}{\thesection\arabic{equation}}
\subsectionfont{\fontsize{12}{14pt plus.8pt minus .6pt}\selectfont}

\usepackage{amsmath}
\usepackage{amssymb}
\usepackage{amsfonts}
\usepackage{multirow}
\usepackage{amsthm}

\newtheorem{theorem}{Theorem}
\newtheorem{lemma}{Lemma}

\theoremstyle{definition}
\newtheorem{definition}{Definition}

\usepackage{url,color} 
\usepackage{algpseudocode}
\usepackage{algorithm}
\usepackage{subfigure}
\usepackage{setspace}
\usepackage{enumitem}
\usepackage{comment}
\usepackage{nomencl} 
\makenomenclature

\usepackage{etoolbox}
\renewcommand\nomgroup[1]{%
	\item[\bfseries
	\ifstrequal{#1}{C}{Important Functions and Constants}{%
		\ifstrequal{#1}{E}{Important Estimators}{}}%
	]}

\let\hat\widehat

\newcommand\R{\mathbb{R}}

\newcommand\E{\mathbb{E}}

\def\young#1{\textcolor[RGB]{152,10,100}{Young: #1}}

\newenvironment{enum}{ 
	\begin{enumerate}
		\setlength{\itemsep}{1pt}
		\setlength{\parskip}{0pt}
		\setlength{\parsep}{0pt}
	}{\end{enumerate}}

\begin{document}


\renewcommand{\baselinestretch}{2}

\markright{ \hbox{\footnotesize\rm 
}\hfill\\[-13pt]
\hbox{\footnotesize\rm
}\hfill }

\markboth{\hfill{\footnotesize\rm YOUNGJUN CHOE, YEN-CHI CHEN, and \textcolor{black}{NICK TERRY}} \hfill}
{\hfill {\footnotesize\rm CROSS-ENTROPY INFORMATION CRITERION} \hfill}

\renewcommand{\thefootnote}{}
$\ $\par


\fontsize{12}{14pt plus.8pt minus .6pt}\selectfont \vspace{0.8pc}
\centerline{\large\bf INFORMATION CRITERION FOR}
\vspace{2pt} \centerline{\large\bf BOLTZMANN APPROXIMATION PROBLEMS}
\vspace{.4cm} \centerline{Youngjun Choe$^{\dagger}$,  Yen-Chi Chen$^{\ddagger}$, and \textcolor{black}{Nick Terry}$^{\dagger}$} \vspace{.4cm} \centerline{\it
$^{\dagger}$Department of Industrial \& Systems Engineering and $^{\ddagger}$Department of Statistics}
\vspace{0cm} \centerline{\it University of Washington, Seattle} \vspace{.55cm} \fontsize{9}{11.5pt plus.8pt minus
.6pt}\selectfont


\begin{quotation}
\noindent {\it Abstract:}
This paper considers the problem of approximating a density when it can be evaluated up to a normalizing constant at a limited number of points. We call this problem the Boltzmann approximation (BA) problem. The BA problem is ubiquitous in statistics, such as approximating a posterior density for Bayesian inference and estimating an optimal density for importance sampling. Approximating the density with a parametric model can be cast as a model selection problem. This problem cannot be addressed with traditional approaches that maximize the (marginal) likelihood of a model, for example, using the Akaike information criterion (AIC) or Bayesian information criterion (BIC). We instead aim to minimize the cross-entropy that gauges the deviation of a parametric model from the target density. We propose a novel information criterion called the cross-entropy information criterion (CIC) and prove that the CIC is an asymptotically unbiased estimator of the cross-entropy (up to a multiplicative constant) under some regularity conditions. We propose an iterative method to approximate the target density by minimizing the CIC. We demonstrate that the proposed method selects a parametric model that well approximates the target density.

\vspace{9pt}
\par
\end{quotation}\par

\def\thefigure{\arabic{figure}}
\def\thetable{\arabic{table}}

\renewcommand{\theequation}{\thesection.\arabic{equation}}

\fontsize{12}{14pt plus.8pt minus .6pt}\selectfont

\setcounter{section}{0} 
\setcounter{equation}{0} 

\lhead[\footnotesize\thepage\fancyplain{}\leftmark]{}\rhead[]{\fancyplain{}\rightmark\footnotesize\thepage}


\section{Introduction}\label{sec:intro}
This paper considers the problem of approximating a target density $q^*(\mathbf{x}) = r(\mathbf{x})/\rho$ when the normalizing constant $\rho$ is unknown but we can evaluate the nonnegative function $r$ at a limited number of points. We call this problem the \emph{Boltzmann approximation problem} (BA problem) due to its relation to the Boltzmann distribution, detailed in Section~\ref{sec:background}. 
This problem is ubiquitous in statistics. Here we list three notable examples where this problem occurs.

{\bf Example 1: Simulation-based inference.}
Stochastic simulation models \citep{Choe2015,pratola2018bayesian} are widely used to estimate
the mean $\mathbb{E}(V)$, where the simulation output of interest, $V$,  depends on the input $\mathbf{X}$ from a known density $p(\mathbf{x})$. 
When the simulator is computationally expensive, the importance sampling estimator $\frac{1}{n}\sum_{i=1}^nV_i\frac{p(\mathbf{X}_i)}{q(\mathbf{X}_i)}$ is widely used because it is 
 unbiased and has the minimum variance when using the output $V_i$ corresponding to the input $\mathbf{X}_i$, $i=1,\ldots,n$, sampled from the optimal density  \citep{chen2019importance}  
$$
q^*(\mathbf{x})\propto \sqrt{\mathbb{E}(V^2|\mathbf{X}=\mathbf{x})} p(\mathbf{x}).
$$
By identifying $r(\mathbf{x}) = \sqrt{\mathbb{E}(V^2|\mathbf{X}=\mathbf{x})} p(\mathbf{x})$, we encounter the BA problem.

{\bf Example 2: Causal inference.}
Consider a simple causal inference problem where we have a response variable $Y$ and a binary treatment $A \in \{0,1\}$. 
Under the potential outcome model \citep{rubin2005causal}, 
the causal effect from $A$ on $Y$ can be analyzed via
studying how the two potential outcomes, $Y(1)$ and $Y(0)$, differ.
To investigate this, we need to know four distributions: 
Two of them, $p(Y(0)=y_0|A=0)$ and $p(Y(1)=y_1|A=1)$, can be directly identified from the data while the other two, $p(Y(0)=y_0|A=1)$ and $p(Y(1)=y_1|A=0)$, cannot.
Luckily, Tukey's factorization shows that 
$$
p(Y(0)=y_0|A=1) \propto\underbrace{\frac{P(A=1|Y(0)=y_0)}{P(A=0|Y(0)=y_0)}}_{O(y_0)} p(Y(0)=y_0|A=0)
$$
and similarly for $p(Y(1)=y_1|A=0)$.
With a model on $O(y_0)$, one approach to approximate $p(Y(0)=y_0|A=1)$ is to sample from its density implied by the above factorization. 
In this case, we only have access to a function that is proportional to the density we want to sample from. 
By identifying $q^*(\cdot) = p(Y(0)=\cdot|A=1)$ and $r(\cdot) = O(\cdot) p(Y(0)=\cdot|A=0)$, we run into the same BA problem.

{\bf Example 3: Bayesian inference.} The most prominent BA problem is approximating a posterior density for Bayesian inference.
Let $\pi(\theta)$ be the prior density of the parameter $\theta$ and $L(\theta|X_1,\cdots,X_n)$ be the likelihood function. 
The posterior (density) is $\pi(\theta|X_1,\cdots, X_n)\propto L(\theta|X_1,\cdots, X_n) \cdot \pi(\theta)$.
In this case, $q^*(\cdot) = \pi(\cdot|X_1,\cdots, X_n)$ is the posterior density and $r(\cdot) = \pi(\cdot)L(\cdot|X_1,\cdots, X_n)$ is the prior density times the likelihood of a model. Then, $\rho$ is the model evidence, which can be compared between models for Bayesian model selection \citep{knuth2015bayesian}. 

When $r$ is computationally light to evaluate, Markov chain Monte Carlo methods have been used extensively in practice, such as the Metropolis-Hastings algorithm that probabilistically accepts or rejects samples \citep{metropolis1953equation,hastings1970monte}, mainly because of theoretical guarantee on asymptotically exact sampling from $q^*$. Also, for exact sampling, the acceptance-rejection sampling algorithm is widely used if an upper-bounding function of $r$ is known.  

But there is growing interest in the BA problems where evaluating $r$ is computationally heavy. In Bayesian inference, $r$ may involve computing the likelihood for a large dataset \citep{Blei2017} or a complex Bayesian model \citep{Beaumont2010,sunnaaker2013approximate}. 
For the causal inference, 
if the model involves covariates (which is often the case), we need to sample counterfactual variables several times per each possible covariate value. 
For simulation-based inference, 
running a simulator is often computationally expensive \citep{Choe2015,pratola2018bayesian}. In such scenarios, one can only afford a relatively small number (e.g., hundreds or thousands) of evaluations of $r$ and does not want to discard any of them.  
This paper focuses on this scenario where we use every evaluation of $r$ to approximate $q^*$. 

While the approximation of $q^*$ is of primary interest, this study also emphasizes the estimation of the normalizing constant $\rho$ because this quantity is, in some applications, even more important to estimate than the density $q^*$ itself. $\rho$ is the model evidence for Bayesian inference and the estimand for importance sampling. In particular, importance sampling requires sampling from the approximate density of $q^*$ and evaluating $r$ at the sampled points to unbiasedly estimate $\rho = \mathbb{E}_{\mu}\!\left[r   \right]$ (with respect to the measure $\mu$) with a minimal variance. 
Therefore, a sampling-based approach to approximate $q^*$ is a key to the BA problem and is the focus of this study to ensure desirable probabilistic guarantees on estimating $\rho$. 

The existing literature on approximating $q^*$ in the BA problem can be generally grouped into parametric approaches \citep{rubinstein1999,Wang2015,Choe2015} and nonparametric approaches \citep{zhang1996nonparametric,neddermeyer2009computationally,chen2019importance}. \citet{chen2019importance} compare both approaches theoretically and empirically. 

This paper considers a class of parametric approaches where we posit a parametric family of densities and find its member closest to $q^*$ in terms of the closeness measured by the Kullback-Leibler (KL) divergence \citep{kullback1951information}. This parametric framework itself is very general and includes maximum likelihood estimation. 

For importance sampling, this framework is first used for the so-called cross-entropy method \citep{rubinstein1999}. 
In this framework, the parametric approximation of $q^*$ takes two steps, namely, 1) choosing a parametric family and 2) minimizing the KL divergence from a member density in the chosen family to the target density $q^*$. The latter is a well-studied optimization problem and the former is an under-explored model selection problem and the focus of this paper.

To tackle this problem, we combine two key ideas: importance sampling
and information criterion. 
Importance sampling \citep{kahn1953} allows us to generate points from distributions different from $q^*$ and still be able to estimate an expectation with respect to $q^*$ up to a constant as long as we have access to $r$. 
With this, we are able to compute the information criterion to perform
parameter estimation and model selection.

To the best of our knowledge, no rigorous framework of combining importance sampling and information criterion is proposed in the literature yet, despite its critical role in choosing an importance sampling density and in  
addressing BA problems. 
We propose a novel information criterion called the cross-entropy information criterion (CIC) and prove that the criterion is an asymptotically unbiased estimator of the KL divergence (up to a multiplicative constant and an additive constant).  We justify that the minimization of the criterion leads to a good model selection.  

We also show that the CIC is reduced to the Akaike information criterion (AIC)\citep{akaike1974} when data are directly generated from the distribution to estimate. Rigorous theoretical analysis of the AIC is a long-standing problem in the literature. Theoretical analyses of information criteria akin to the AIC often impose uniform integrability conditions directly on estimators to make the model complexity penalty to be expressed in the free parameter dimension $d$ (see Conditions A7--A8 in \citet{donohue2011conditional}, Theorem~1 in \citet{Claeskens2008}, and more references cited in \citet[][p.1157]{bhansali1991convergence} and \citet[][p.416]{findley2002aic}). For autoregressive models, sufficient conditions for such uniform integrability conditions are established much later than the seminal paper of the AIC \citep{akaike1974} where the uniform integrability is not established rigorously \citep{bhansali1986derivation,findley2002aic}. For general parametric models, establishing general versions of the sufficient conditions has been an open problem. This paper establishes such sufficient conditions to validate the CIC, which is more general than AIC.


The remainder of this paper is organized as follows. Section~\ref{sec:background} briefly reviews the relevant background. Section~\ref{sec:CIC} proposes the cross-entropy information criterion. Section~\ref{sec:application_CIC} explains how the proposed criterion can be used in practice 
with a numerical example 
for approximating an optimal density for importance sampling. Section~\ref{sec:conclusion} concludes the paper. 



%



\section{Background}\label{sec:background}
This section briefly explains the origin of the BA problem and reviews the KL divergence, the maximum likelihood estimator (MLE), and the Akaike information criterion (AIC) to a) introduce the minimum cross-entropy estimator (MCE), which is a generalized version of MLE, 
and b) pave the way for generalizing the AIC to the cross-entropy information criterion (CIC). 

The name of the BA problem originates from statistical physics where 
the nonnegative function
$r(\mathbf{x})$ is what we can evaluate 
and 
express as 
$
r(\mathbf{x}) = e^{-\phi(\mathbf{x})}, 
$
where $\phi(\mathbf{x})$ is often called the energy of a state $\mathbf{x}$ that can be calculated using theories in physics. 
Using this notation, the target density $q^*(\mathbf{x}) \propto e^{-\phi(\mathbf{x})}$
is called the Boltzmann distribution
and $\rho = \int r(\mathbf{x})\,\mathrm{d}\mathbf{x} = \int e^{-\phi(\mathbf{x})} \,\mathrm{d}\mathbf{x}$
is called the partition function. 
The BA problem often has one practical constraint: evaluating $r$ is (computationally) expensive so we want to make good use of all evaluations. 
The KL divergence \citep{kullback1951information} is commonly used to gauge the difference between two distributions in statistical inference. Consider two probability measures, $Q^*$ and $Q$, on a common measurable space 
such that $Q^*$ is absolutely continuous with respect to $Q$ (written $Q^* \ll Q$). Then, the KL divergence from $Q$ to $Q^*$ is defined as 
\begin{align}
\mathbb{D}\!\left(Q^*|| Q\right) &:= \mathbb{E}_{Q^*} \!\left[\log{\frac{\mathrm{d}Q^*}{\mathrm{d}Q}}\right] , \label{eq:KL_divergence}
\end{align}
where $\mathbb{E}_{Q^*}$ denotes the expectation with respect to $Q^*$ and ${\mathrm{d}Q^*}/{\mathrm{d}Q}$ is the Radon–-Nikodym derivative of $Q^*$ with respect to $Q$. If $Q^*$ and $Q$ are absolutely continuous with respect to a dominating measure $\mu$ (e.g., counting or Lebesgue measure), their respective densities, $q^*$ and $q$, exist by the Radon–-Nikodym theorem and  the KL divergence in \eqref{eq:KL_divergence} can be expressed as
\begin{align}
\mathbb{D}\!\left(Q^*|| Q\right) &= \mathbb{E}_{Q^*} \!\left[\log{\frac{q^*}{q}}\right]  \nonumber
\\&= \int q^* \log{\left(\frac{q^*}{q}\right)}  \,\mathrm{d}\mu . \nonumber
\end{align}
The KL divergence is non-negative and takes zero if and only if $q^* = q$ almost everywhere (a.e.). 
If $Q$ is an approximation of $Q^*$, the choice of $Q$ minimizing $\mathbb{D}\!\left(Q^*|| Q\right)$ leads to a good approximation.

The MLE is a prominent example of using the KL divergence. When the data $\mathbf{X}_1, \ldots, \mathbf{X}_n$ are drawn from an unknown distribution $Q^*$, we can approximate $Q^*$ by a distribution in a parametric family $\{Q_{\boldsymbol{\theta}}: \boldsymbol{\theta} \in \boldsymbol{\Theta}_d \subset \mathbb{R}^d\}$ by minimizing the KL divergence from $Q_{\boldsymbol{\theta}}$ to $Q^*$ over $\boldsymbol{\theta} \in \boldsymbol{\Theta}_d$. Suppose $Q^* \ll Q_{\boldsymbol{\theta}}$ for all $\boldsymbol{\theta} \in \boldsymbol{\Theta}_d$ so that the KL divergence is well defined over $\boldsymbol{\Theta}_d$. Also, suppose $Q_{\boldsymbol{\theta}} \ll \mu$ for all $\boldsymbol{\theta} \in \boldsymbol{\Theta}_d$ and $Q^* \ll \mu$ so that densities $q_{\boldsymbol{\theta}} = {\mathrm{d}Q_{\boldsymbol{\theta}}}/{\mathrm{d}\mu}$ and $q^*={\mathrm{d}Q^*}/{\mathrm{d}\mu}$ exist. Then, the KL divergence is
\begin{align}
\mathbb{D}\!\left(Q^*|| Q_{\boldsymbol{\theta}}\right) &= \int q^* \log{\left(\frac{q^*}{q_{\boldsymbol{\theta}}}\right)}  \,\mathrm{d}\mu  \nonumber
\\&=  \int q^* \log{q^*} \,\mathrm{d}\mu - \int q^* \log{q_{\boldsymbol{\theta}}} \,\mathrm{d}\mu . \label{eq:KL_divergence2} 
\end{align}
Note that only the second term in \eqref{eq:KL_divergence2}, called cross-entropy, depends on $\boldsymbol{\theta}$. Therefore, minimizing the KL divergence over $\boldsymbol{\theta} \in \boldsymbol{\Theta}_d$ is equivalent to minimizing the cross-entropy over $\boldsymbol{\theta} \in \boldsymbol{\Theta}_d$. Because $q^*$ is unknown, the cross-entropy should be estimated based on $\mathbf{X}_1, \ldots, \mathbf{X}_n \sim Q^*$. An unbiased, consistent estimator of the cross-entropy is
\begin{align}
- \frac{1}{n}\sum_{i=1}^{n}  \log{q_{\boldsymbol{\theta}}\!\left(\mathbf{X}_i\right)}  ,  \label{eq:avg_likelihood} 
\end{align}
which is the average of negative log-likelihoods for the observed data. Therefore, the MLE of $\boldsymbol{\theta}$, denoted by $\boldsymbol{\hat{\theta}}_n$, is the minimizer of the cross-entropy estimator in \eqref{eq:avg_likelihood}. 

Another example of using the KL divergence or cross-entropy is the Akaike information criterion (AIC) \citep{akaike1974}. As the free parameter dimension $d$ of the parameter space $\boldsymbol{\Theta}_d$ (or equivalently, the model degrees of freedom) increases, $Q_{\boldsymbol{\hat{\theta}}_n}$ may become a better approximation of $Q^*$. To compare the different approximating distributions (or models), we could use a plug-in estimator 
\begin{align}
- \frac{1}{n}\sum_{i=1}^{n}  \log{q_{\boldsymbol{\hat{\theta}}_n}\!\left(\mathbf{X}_i\right)}    \label{eq:biased_avg_likeli}
\end{align}
of the cross-entropy, but this is problematic because of the downward bias created from using the data twice (once for $\boldsymbol{\hat{\theta}}_n$ and another for estimating the cross-entropy). 
While complex models (or distributions with larger $d$'s) tend to have a smaller cross-entropy estimate, they are subject to the overfitting problem. The AIC remedies this issue by correcting the asymptotic bias of the estimator in \eqref{eq:biased_avg_likeli}. The AIC is defined (up to a multiplicative constant) as 
\begin{align}
- \frac{1}{n}\sum_{i=1}^{n}  \log{q_{\boldsymbol{\hat{\theta}}_n}\!\left(\mathbf{X}_i\right)} + \frac{d}{n} ,  \label{eq:AIC}
\end{align}
where the bias correction term $d/n$ penalizes the model complexity, balancing it with the goodness-of-fit represented by the first term. Minimizing the AIC can be interpreted as minimizing an asymptotically unbiased estimator of the cross-entropy. Therefore, both the MLE and AIC aim at minimizing the cross-entropy from an approximate distribution $Q_{\boldsymbol{\theta}}$ to the unknown distribution $Q^*$ generating the data.


\section{Cross-Entropy Information Criterion for the BA problem}\label{sec:CIC}

This section develops an information criterion for the BA problem. 
We introduce an approximation of the cross-entropy 
in the BA problem that is suitable for parameter estimation
and an information criterion that can be used for model selection. 


\subsection{Approximate Cross-Entropy for the BA problem}


Analogous to the MLE and AIC, our approximation task in this paper considers minimizing the KL divergence (or cross-entropy) from a parametric distribution $Q_{\boldsymbol{\theta}}$ to the target distribution $Q^*$ over $\boldsymbol{\theta} \in \boldsymbol{\Theta}_d$ (to well-define the KL divergence, hereafter assume $Q^* \ll Q_{\boldsymbol{\theta}}$ for all $\boldsymbol{\theta} \in \boldsymbol{\Theta}_d$) when the target density $q^*$ is proportional to a nonnegative function $r$, i.e., $q^* =  r / \rho$ for a positive unknown constant $\rho = \int r  \,\mathrm{d}\mu$.
Minimizing the KL divergence in \eqref{eq:KL_divergence2} over $\boldsymbol{\theta}$ is equivalent to minimizing the cross-entropy 
$
-\int q^*\log q_{\boldsymbol{\theta}} \,\mathrm{d}\mu =-\frac{1}{\rho} \int r\log q_{\boldsymbol{\theta}} \,\mathrm{d}\mu
$
and 
equivalent to minimizing 
\begin{align}
\mathcal{C}(\boldsymbol{\theta}) &:= - \int r \log{q_{\boldsymbol{\theta}}} \,\mathrm{d}\mu  ,  \label{eq:C_theta}
\end{align}
which is unknown in practice because $r$ can be evaluated only at observed data points. Using the idea of importance sampling, we approximate $ \mathcal{C}(\boldsymbol{\theta})$ in \eqref{eq:C_theta} by 
an unbiased and consistent estimator
\begin{align}
\bar{\mathcal{C}}_{\boldsymbol{\eta}}(\boldsymbol{\theta}) &:= - \frac{1}{n}\sum_{i=1}^{n} \frac{r(\mathbf{X}_i)}{q_{\boldsymbol{\eta}}\!\left(\mathbf{X}_i \right)}   \log{q_{\boldsymbol{\theta}}\!\left(\mathbf{X}_i \right)}  \label{eq:C_bar_theta}
\end{align}
where $\mathbf{X}_1, \ldots, \mathbf{X}_n \sim Q_{\boldsymbol{\eta}}$ with any parameter $\boldsymbol{\eta} \in \boldsymbol{\Theta}_d$ as long as $q_{\boldsymbol{\eta}}$ covers the support of $q^*$.
Because the estimator in \eqref{eq:C_bar_theta} is to approximate the cross-entropy
in the BA problem,
we call it the \emph{approximate cross-entropy} (ACE).

Therefore, by minimizing $\bar{\mathcal{C}}_{\boldsymbol{\eta}}(\boldsymbol{\theta})$ in \eqref{eq:C_bar_theta} over $\boldsymbol{\theta} \in \boldsymbol{\Theta}_d$, we can approximately minimize the KL divergence (or cross-entropy) from $Q_{\boldsymbol{\theta}}$ to $Q^*$. Thus, we call
\begin{align}
\boldsymbol{\hat{\theta}}_n := \underset{\boldsymbol{\theta}\in \boldsymbol{\Theta}_d}{\arg\!\min}\,  \bar{\mathcal{C}}_{\boldsymbol{\eta}}(\boldsymbol{\theta}) \label{eq:MCE}
\end{align}
the \emph{minimum cross-entropy estimator} (MCE) because
it minimizes the ACE. 
Note that if the random sample is directly drawn from the target distribution (i.e., $\mathbf{X}_1, \ldots, \mathbf{X}_n \sim Q_{\boldsymbol{\eta}}=Q^*$), then the MCE reduces to the MLE because minimizing \eqref{eq:C_bar_theta} is equivalent to minimizing \eqref{eq:avg_likelihood} due to $r / q^* = \rho$. 

In the importance sampling literature,
the density $q_{\boldsymbol{\eta}}$ in \eqref{eq:C_bar_theta} is called the importance sampling (or proposal) density and $q^*$ the optimal importance sampling density. The density $q^*$ is optimal because 
the variance of the following importance sampling estimator of $\rho$ is reduced to zero if $\mathbf{X}_1, \ldots, \mathbf{X}_n$ are drawn from $q_{\boldsymbol{\eta}}=q^*$:
\begin{align}
\hat{\rho}_{\textrm{IS}} &= \frac{1}{n}\sum_{i=1}^{n} \frac{r\!\left(\mathbf{X}_i\right)}{q_{\boldsymbol{\eta}}\!\left(\mathbf{X}_i \right)} .     \label{eq:rho_IS}
\end{align}
In practice, $q^*$ is unknown and thus approximated by $q_{\boldsymbol{\eta}}$. Therefore, finding $q_{\boldsymbol{\eta}}$ closest to $q^*$ is of primary interest. 

\subsection{Properties of the MCE}
The MCE is a minimizer of \eqref{eq:C_bar_theta} so it is an M-estimator \citep{huber1964robust}, which allows us to use the M-estimation theory \citep{van1998asymptotic}. Hereafter, we assume the parameter minimizing the true cross-entropy, $$\boldsymbol{\theta}^* := \underset{\boldsymbol{\theta}\in \boldsymbol{\Theta}_d}{\arg\!\min}\, \mathcal{C}(\boldsymbol{\theta}),$$ is unique. Then, the MCE $\boldsymbol{\hat{\theta}}_n$ is a strongly consistent estimator of $\boldsymbol{\theta}^*$ (Lemma~\ref{lemma:consistency}) and has asymptotic normality (Lemma~\ref{lemma:asymp_norm}) under standard regularity conditions (see assumptions in Appendix A).   


\begin{lemma}[Strong consistency of the MCE] \label{lemma:consistency}
	Suppose that assumptions (A1) and (B1-2)
	hold and $\boldsymbol{\Theta}_d$ is compact. Then, for any $\boldsymbol{\eta} \in \boldsymbol{\Theta}_d$,  
	the MCE 
	$\boldsymbol{\hat{\theta}}_n := \underset{\boldsymbol{\theta}\in \boldsymbol{\Theta}_d}{\arg\!\min}\, \bar{\mathcal{C}}_{\boldsymbol{\eta}}(\boldsymbol{\theta})$
	converges almost surely to $\boldsymbol{\theta}^*$ as $n\to\infty$.
\end{lemma}

\begin{lemma}[Asymptotic normality of the MCE] \label{lemma:asymp_norm}
	Suppose that assumptions (A1-3) and (B3-4)
	hold and that for any $\boldsymbol{\eta} \in \boldsymbol{\Theta}_d$, $\boldsymbol{\hat{\theta}}_n$ 
	converges in probability to $\boldsymbol{\theta}^*$ as $n\to\infty$. 
	Then, for any $\boldsymbol{\eta} \in \boldsymbol{\Theta}_d$,  $\sqrt{n}\left(\boldsymbol{\hat{\theta}}_n-\boldsymbol{\theta}^*\right)$ converges in distribution to $N\!\left(0, \boldsymbol{\Gamma}^{-1}\boldsymbol{\Lambda}_{\boldsymbol{\eta}}\boldsymbol{\Gamma}^{-1} \right)$ as $n\to\infty$,  where 
	$\boldsymbol{\Gamma} :=-\mathbb{E}_{\mu}\!\left[r  \nabla_{\boldsymbol{\theta}}^2 \log{q_{\boldsymbol{\theta}^*}} \right] \nonumber$
	and 
	$\boldsymbol{\Lambda}_{\boldsymbol{\eta}} := \mathbb{E}_{Q_{\boldsymbol{\eta}}}\!\left[\nabla_{\boldsymbol{\theta}} h(\mathbf{X}, \boldsymbol{\eta}, \boldsymbol{\theta}^*)  \nabla_{\boldsymbol{\theta}} h(\mathbf{X}, \boldsymbol{\eta}, \boldsymbol{\theta}^*)^T \right]   \nonumber$
	with 
	\begin{align}
	h\!\left(\mathbf{X}, \boldsymbol{\eta}, \boldsymbol{\theta}\right) := \frac{r(\mathbf{X})}{q_{\boldsymbol{\eta}}\!\left(\mathbf{X} \right)}   \log{q_{\boldsymbol{\theta}}\!\left(\mathbf{X} \right)} .  \label{eq:func_h}
	\end{align}
\end{lemma}

The above two lemmas are common results from the standard assumptions, see, e.g., Theorem A1 and A2 in \citet{rubinstein1993}.

\subsection{Iterative procedure for minimizing the ACE}

Although the ACE $\bar{\mathcal{C}}_{\boldsymbol{\eta}}(\boldsymbol{\theta})$ in \eqref{eq:C_bar_theta} is an unbiased estimator of the cross-entropy (up to a multiplicative constant) from $Q_{\boldsymbol{\theta}}$ to $Q^*$, 
a bad choice of sampling density $q_{\boldsymbol{\eta}}$
may lead to a large variance. 
Inspired by the cross-entropy method \citep{rubinstein1999},
we propose to update  $\boldsymbol\eta$ iteratively by the closest
estimate of $\boldsymbol\theta^*$ as described in the procedure in Figure~\ref{fig:MCE_iter}. We use the same sample size $n$ for each iteration for notational simplicity without loss of generality. 

\begin{figure}[!h]
	\fbox{\parbox{5.9in}{
			\begin{center}
				{\sc Iterative procedure for approximating $Q^*$}
			\end{center}
			Inputs: iteration counter $t=1$, the number of iterations $\tau$, the sample size $n$, and the initial parameter $\boldsymbol{\hat{\theta}}_n^{(0)} = \boldsymbol{\eta} \in \boldsymbol{\Theta}_d$.
			\begin{center}
				\begin{enum}
					\item[1.] Sample $\mathbf{X}_{1}^{(t-1)},\ldots, \mathbf{X}_{n}^{(t-1)}$ from $Q_{\boldsymbol{\hat{\theta}}_n^{(t-1)}}$. 
					\item[2.] Find the MCE $\boldsymbol{\hat{\theta}}_n^{(t)}:= \underset{\boldsymbol{\theta}\in \boldsymbol{\Theta}_d}{\arg\!\min}\, \bar{\mathcal{C}}_{\boldsymbol{\hat{\theta}}_n^{(t-1)}}(\boldsymbol{\theta})$, where 
					\begin{align}
					\bar{\mathcal{C}}_{\boldsymbol{\hat{\theta}}_n^{(t-1)}}(\boldsymbol{\theta}) &:= - \frac{1}{n}\sum_{i=1}^{n} \frac{r\!\left(\mathbf{X}_i^{(t-1)}\right)}{q_{\boldsymbol{\hat{\theta}}_n^{(t-1)}}\!\left(\mathbf{X}_i^{(t-1)} \right)}   \log{q_{\boldsymbol{\theta}}\!\left(\mathbf{X}_i^{(t-1)} \right)}, \nonumber
					\\&= -\frac{1}{n  }	\sum_{i=1}^{n}	h\!\left(\mathbf{X}_i^{(t-1)}, \boldsymbol{\hat{\theta}}_n^{(t-1)}, \boldsymbol{\theta}\right) \label{eq:C-bar_iter}   
					\end{align}
					\item[3.] If $t = \tau$, output the approximate distribution  $Q_{\boldsymbol{\hat{\theta}}_n^{(\tau)}}$.  Otherwise, increment $t$ by 1 and go to Step 1.
				\end{enum}
			\end{center}
	}}
	\caption{Iterative procedure for approximating $Q^*$ by minimizing the estimator of cross-entropy from a parametric distribution $Q_{\boldsymbol{\theta}}$ to $Q^*$.}
	\label{fig:MCE_iter}
\end{figure}

An important property of the procedure in Figure~\ref{fig:MCE_iter} is that
the iterative update of $\boldsymbol\eta$ by the best estimate of $\boldsymbol\theta^*$ leads to an information criterion that
mimics the AIC
under good regularity conditions. 
If $\boldsymbol\eta$ is fixed, we will not obtain
such a nice property.

The iterative approach can be used to estimate $\rho$, the normalizing constant,
by modifying the estimator in \eqref{eq:rho_IS} as follows:
\begin{align}
\hat{\rho}^{(t-1)} &= \frac{1}{n}\sum_{i=1}^{n} \frac{r\!\left(\mathbf{X}_i^{(t-1)}\right)}{q_{\boldsymbol{\hat{\theta}}_n^{(t-1)}}\!\left(\mathbf{X}_i^{(t-1)} \right)}     \label{eq:rho^t-1}.
\end{align}
Furthermore,  if $q_{\boldsymbol{\hat{\theta}}_n^{(t-1)}} = q^*$, the estimator in \eqref{eq:rho^t-1} is the optimal importance sampling estimator having zero variance \citep{kahn1953}. Because the iterative procedure refines $q_{\boldsymbol{\hat{\theta}}_n^{(t-1)}}$ to be closer to $q^*$, $\hat{\rho}^{(t-1)}$ will generally have a smaller variance as $t$ gets larger. 

The fact that we can estimate $\rho$ with a small variance as a byproduct of approximating $q^*$ is particularly desirable, because $\rho$, the normalizing constant of $q^* =  r / \rho$, is often a quantity of interest as discussed in Section~\ref{sec:intro}.



\subsection{Approximating AIC}\label{subsec:asymp_bias}

To simplify the model complexity penalty term in the AIC, \citet{akaike1974} assume that the true data generating distribution belongs to the parametric distribution family being considered. 
We make a similar assumption $Q^* = Q_{\boldsymbol{\theta}^*}$ (i.e., assumption (A1) in Appendix A)
 to simplify the asymptotic bias of $\bar{\mathcal{C}}_{\boldsymbol{\hat{\theta}}_n^{(t-1)}}\!\left(\boldsymbol{\hat{\theta}}_n^{(t)}\right)$ in estimating $\mathcal{C}\!\left(\boldsymbol{\hat{\theta}}_n^{(t)}\right)$.


In what follows we describe
our main result as Theorem~\ref{thm:asymp_bias} that quantifies the asymptotic bias.
Before deriving the asymptotic bias, we first introduce
a useful lemma that
characterizes the limiting behavior 
of any MCE.
The assumptions and proofs are deferred to
Appendix~A.
We note that as $n$ tends to infinity, the number of iterations $\tau \ge 2$ remains fixed and the 
parameter dimension $d$ is also fixed.



\begin{lemma}\label{lem:expectation}
	Suppose that assumptions (A1-5) and (B1-5) hold
	and let $\hat {\boldsymbol{\theta}}_n$
	be the MCE. 
	Then 
	$
	\sqrt{n}\left(\boldsymbol{\hat{\theta}}_n-\boldsymbol{\theta}^*\right) = Z_{\boldsymbol{\eta}} + \epsilon_{n,\boldsymbol{\eta}}$
	where 
	\begin{align*}
	Z_{\boldsymbol{\eta}}:=-\boldsymbol{\Gamma}^{-1}\frac{1}{\sqrt{n}}\sum_{i=1}^n\nabla_{\boldsymbol{\theta}} h(\mathbf{X}_i;\boldsymbol{\eta},\boldsymbol{\theta}^*),\quad
	\sup_{\boldsymbol{\eta}\in \boldsymbol{\Theta}_d}\E(\|\epsilon_{n,\boldsymbol{\eta}}\|) = O(1/n^{1/4})
	\end{align*}
	and
	$\E(Z_{\boldsymbol{\eta}}) = 0,\quad
	{\sf Cov}(Z_{\boldsymbol{\eta}})= \boldsymbol{\Gamma}^{-1}\boldsymbol{\Lambda}_{\boldsymbol{\eta}}\boldsymbol{\Gamma}^{-1},\quad
	\boldsymbol{\Lambda}_{\boldsymbol{\eta}} = \rho \boldsymbol{\Gamma} +  O(\|\boldsymbol{\eta} - \boldsymbol{\theta}^*\|).$
\end{lemma}

Lemma~\ref{lem:expectation} is different from
the conventional asymptotic normality (e.g., Lemma~\ref{lemma:asymp_norm})
in two senses. 
First, we require a bound on the expectation of the remainder term $\E(\|\epsilon_{n,\boldsymbol{\eta}}\|)$,
which implies the rate of $O_P(1/n^{1/4})$
and the asymptotic normality. 
We need the expectation because the derivation of the asymptotic bias
in the CIC analysis requires an expectation bound. 
Second, 
the expectation of the remainder term is bounded uniformly for all $\boldsymbol{\eta}\in\boldsymbol{\Theta}_d$.
This is in contrast to the conventional analysis that only requires the result at $\boldsymbol{\eta}=\boldsymbol{\theta}^*$.
We need the uniform bound over $\boldsymbol{\eta}\in\boldsymbol{\Theta}_d$ because 
the data are generated from a measure $Q_{\boldsymbol{\eta}}$ where $\boldsymbol{\eta}$
is a random vector.
In our analysis,
we generate a sequence of estimators $\hat{\boldsymbol{\theta}}_n^{(1)},\hat{\boldsymbol{\theta}}_n^{(2)},\cdots,\hat{\boldsymbol{\theta}}_n^{(\tau)}$
and the data used to estimate $\hat{\boldsymbol{\theta}}_n^{(t)}$
are generated from $Q_{\hat{\boldsymbol{\theta}}_n^{(t-1)}}$,
the measure based on the previous set of data.
The uniform bound ensures that the remainder term $\epsilon_{n,\boldsymbol{\eta}}$
is small even if $\boldsymbol{\eta}$ is a random quantity.
For any sequential/iterative sampling procedure, 
we would need a similar bound to control
the remainder terms.



\begin{theorem}[Asymptotic bias of $\bar{\mathcal{C}}_{\boldsymbol{\hat{\theta}}_n^{(t-1)}}\!\left(\boldsymbol{\hat{\theta}}_n^{(t)}\right)$ in estimating $\mathcal{C}\!\left(\boldsymbol{\hat{\theta}}_n^{(t)}\right)$]\label{thm:asymp_bias}
	Suppose that assumptions (A1-6) and (B1-5) hold.
	Then 	
	\begin{align}
	\mathbb{E}\!\left[\bar{\mathcal{C}}_{\boldsymbol{\hat{\theta}}_n^{(t-1)}}\!\left(\boldsymbol{\hat{\theta}}_n^{(t)}\right)  - \mathcal{C}\!\left(\boldsymbol{\hat{\theta}}_n^{(t)}\right) \right] &= -\rho\frac{d}{n} + o\!\left(\frac{1}{n}\right) \nonumber 
	\end{align}
	for each $t=2,\ldots,\tau$.
\end{theorem}

The asymptotic bias, $-\rho{d}/{n}$, is proportional to the free parameter dimension $d$ of the parameter space $\boldsymbol{\Theta}_d$, similar to the penalty term of the AIC in \eqref{eq:AIC}. 
In practice, 
$\rho = \mathbb{E}_{\mu}\!\left[r   \right]$ is unknown, but we can use a consistent estimator of $\rho$ to estimate the asymptotic bias, such as the estimator in \eqref{eq:rho^t-1}.


As a bias-corrected estimator of the cross-entropy $\mathcal{C}\!\left(\boldsymbol{\hat{\theta}}_n^{(t)}\right)$ (up to a multiplicative constant), we 
define the \emph{cross-entropy information criterion} (CIC) as the follows:
\\
\fbox{\parbox{5.9in}{
		\begin{definition}[Cross-entropy information criterion (CIC)] 
			\begin{align}
			\mathrm{CIC}^{(t)}(d) = \bar{\mathcal{C}}_{\boldsymbol{\hat{\theta}}_n^{(t-1)}}\!\left(\boldsymbol{\hat{\theta}}_n^{(t)}\right) + \hat{\rho}\frac{d}{n} \label{eq:CIC_uncumul}
			\end{align}
			for $t=1,\ldots,\tau$
			, where $\boldsymbol{\hat{\theta}}_n^{(t)} := \underset{\boldsymbol{\theta}\in \boldsymbol{\Theta}_d}{\arg\!\min}\, \bar{\mathcal{C}}_{\boldsymbol{\hat{\theta}}_n^{(t-1)}}(\boldsymbol{\theta})$ with $\bar{\mathcal{C}}_{\boldsymbol{\hat{\theta}}_n^{(t-1)}}(\cdot)$ in \eqref{eq:C-bar_iter}. $\hat{\rho}$ is a consistent estimator of $\rho$, such as the estimator in \eqref{eq:rho^t-1}.
\end{definition}}}


We note that the CIC reduces to the AIC up to an additive $o_p(1/n)$ if the samples are all drawn from the target distribution, that is, $\mathbf{X}_{1}^{(t-1)},\ldots, \mathbf{X}_{n}^{(t-1)} \sim Q_{\boldsymbol{\hat{\theta}}_n^{(t-1)}} = Q^*$ for $t=1,\ldots,\tau$ in Figure~\ref{fig:MCE_iter}. If so, the first term of the CIC in \eqref{eq:CIC_uncumul} becomes
\begin{align}
\bar{\mathcal{C}}_{\boldsymbol{\hat{\theta}}_n^{(t-1)}}\!\left(\boldsymbol{\hat{\theta}}_n^{(t)}\right) &:= - \frac{1}{n}\sum_{i=1}^{n} \frac{r\!\left(\mathbf{X}_i^{(t-1)}\right)}{q_{\boldsymbol{\hat{\theta}}_n^{(t-1)}}\!\left(\mathbf{X}_i^{(t-1)} \right)}   \log{q_{\boldsymbol{\hat{\theta}}_n^{(t)}}\!\left(\mathbf{X}_i^{(t-1)} \right)} \label{eq:CIC-first-line1}
\\&= - \frac{1}{n}\sum_{i=1}^{n} \frac{r\!\left(\mathbf{X}_i^{(t-1)}\right)}{q^*\!\left(\mathbf{X}_i^{(t-1)} \right)}   \log{q_{\boldsymbol{\hat{\theta}}_n^{(t)}}\!\left(\mathbf{X}_i^{(t-1)} \right)} \label{eq:CIC-first-line2}
\\&= - \frac{\rho}{n}\sum_{i=1}^{n}  \log{q_{\boldsymbol{\hat{\theta}}_n^{(t)}}\!\left(\mathbf{X}_i^{(t-1)} \right)} , \label{eq:CIC-first-line3}
\end{align}
because $q_{\boldsymbol{\hat{\theta}}_n^{(t-1)}} = q^*$ in \eqref{eq:CIC-first-line1} and $r / q^* = \rho$ in \eqref{eq:CIC-first-line2}. Plugging the expression in \eqref{eq:CIC-first-line3} into the CIC in \eqref{eq:CIC_uncumul} shows that the CIC is equal to $\rho$ times the AIC in \eqref{eq:AIC} up to an additive $o_p(1/n)$.  

The asymptotic bias expression in Theorem~\ref{thm:asymp_bias} holds only for $t \ge 2$, because when $t=1$, the initial sample is drawn from $Q_{\boldsymbol{\hat{\theta}}^{(0)}_n} = Q_{\boldsymbol{\eta}}$, which is not a distribution converging to $Q_{\boldsymbol{\theta}^*}$. 
If we want to select a reasonable parameter dimension $d$ to use at the first iteration, it is still necessary to penalize the model complexity. 
Therefore, we define the CIC even for $t=1$. 


\subsection{The CIC based on cumulative data}
If we use the equal sample size $n$ for each iteration, the model dimension $d$ for later iterations may vary only a little from the earlier iterations. Alternatively, 
we can aggregate the samples gathered through iterations to obtain a cumulative version of the CIC as discussed in this subsection. 

In the $t^{th}$ iteration, the cumulative version uses all the observed data up to the current iteration to estimate $\mathcal{C}(\boldsymbol{\theta})$, instead of using only the current iteration's data $\mathbf{X}_{1}^{(t-1)},\ldots, \mathbf{X}_{n}^{(t-1)} \sim Q_{\boldsymbol{\hat{\theta}}_n^{(t-1)}}$ (recall Figure~\ref{fig:MCE_iter}).  
The benefit of the cumulative version is the tendency of the aggregated estimator of $\mathcal{C}(\boldsymbol{\theta})$ to have a smaller variance than the non-aggregated estimator $\bar{\mathcal{C}}_{\boldsymbol{\hat{\theta}}_n^{(t-1)}}(\boldsymbol{\theta})$ in \eqref{eq:C-bar_iter}. 
This approach, in turn, can reduce the variance of the MCE as well, which minimizes the aggregated estimator of $\mathcal{C}(\boldsymbol{\theta})$. 

For more flexibility, we can allocate a different sample size for each iteration, that is, $n_t$ for the $t^{th}$ iteration, $t=0,1,\ldots,\tau$ (for example, a large $n_0$ for the initial sample to broadly cover the support of $Q_{\boldsymbol{\eta}}$ and equal sample sizes $n_1 = \ldots = n_\tau$ for the later iterations). 
Then, we can  find the MCE
\begin{align}
\boldsymbol{\hat{\theta}}^{(t)} &:= \underset{\boldsymbol{\theta}\in \boldsymbol{\Theta}_d}{\arg\!\min}\, \bar{\mathcal{C}}^{(t-1)}(\boldsymbol{\theta}) , \label{eq:MCE_cumul}
\end{align}
where the aggregated estimator of $\mathcal{C}(\boldsymbol{\theta})$ is denoted as  
\begin{align}
\bar{\mathcal{C}}^{(t-1)}(\boldsymbol{\theta}) &:= \frac{1}{\sum_{s = 0}^{t-1}n_{s}}\sum_{s = 0}^{t-1} n_{s} \bar{\mathcal{C}}_{\boldsymbol{\hat{\theta}}^{(s)}}(\boldsymbol{\theta})   \nonumber
\\&=  -\frac{1}{\sum_{s = 0}^{t-1}n_{s}}  \sum_{s = 0}^{t-1}    \sum_{i=1}^{n_s} \frac{r\!\left(\mathbf{X}_i^{(s)}\right)}{q_{\boldsymbol{\hat{\theta}}^{(s)}}\!\left(\mathbf{X}_i^{(s)} \right)}   \log{q_{\boldsymbol{\theta}}\!\left(\mathbf{X}_i^{(s)} \right)}  \nonumber
\\&= -\frac{1}{\sum_{s = 0}^{t-1}n_{s}}\sum_{s = 0}^{t-1} 	\sum_{i=1}^{n_s}	h\!\left(\mathbf{X}_i^{(s)}, \boldsymbol{\hat{\theta}}^{(s)}, \boldsymbol{\theta}\right)  \label{eq:CE_cumulative}
\end{align}
for $t = 1,\ldots, \tau$  with $\boldsymbol{\hat{\theta}}^{(0)} := \boldsymbol{\eta}$.  We note that $\bar{\mathcal{C}}^{(t-1)}(\boldsymbol{\theta})$ in \eqref{eq:CE_cumulative} is an unbiased estimator of $\mathcal{C}(\boldsymbol{\theta})$. 

By using all data gathered up to the $t^{\textrm{th}}$ iteration, we can determine the model parameter dimension $d$ at the $t^{th}$ iteration with the following CIC:
\\
\fbox{\parbox{5.9in}{
		\begin{definition}[CIC: Cumulative version]
			\begin{align}
			\overline{\mathrm{CIC}}^{(t)}(d) = \bar{\mathcal{C}}^{(t-1)}\!\left(\boldsymbol{\hat{\theta}}^{(t)}\right) + \hat{\rho}\frac{d}{\sum_{s = 0}^{t-1}n_{s}}  \label{eq:CIC_cumul}
			\end{align}
			for $t=1,\ldots,\tau$
			, where $\boldsymbol{\hat{\theta}}^{(t)} := \underset{\boldsymbol{\theta}\in \boldsymbol{\Theta}_d}{\arg\!\min}\, \bar{\mathcal{C}}^{(t-1)}(\boldsymbol{\theta})$ in \eqref{eq:MCE_cumul}. $\hat{\rho}$ is a consistent estimator of $\rho$, such as the estimators in \eqref{eq:rho_hat_cumul0} and \eqref{eq:rho_hat_cumul}.
\end{definition}}}

\noindent As $t$ increases, the accumulated sample size $\sum_{s = 0}^{t-1}n_{s}$ increases so that the free parameter dimension $d$ can increase. Thus, the cumulative version of the CIC allows the use of a highly complex model if it can better approximate $Q^*$. 




As a consistent and unbiased estimator of $\rho$, we can use 
\begin{align}
\hat{\rho}^{(0)} &= \frac{1}{n_{0}}    \sum_{i=1}^{n_0} \frac{r\!\left(\mathbf{X}_i^{(0)}\right)}{q_{\boldsymbol{\hat{\theta}}^{(0)}}\!\left(\mathbf{X}_i^{(0)} \right)} \label{eq:rho_hat_cumul0}
\end{align}
at the $1^{st}$ iteration. At the $t^{th}$ iteration for $t=2,\ldots,\tau$, we can use
\begin{align}
\hat{\rho}^{(t-1)} &= \frac{1}{\sum_{s = 1}^{t-1}n_{s}}  \sum_{s = 1}^{t-1}    \sum_{i=1}^{n_s} \frac{r\!\left(\mathbf{X}_i^{(s)}\right)}{q_{\boldsymbol{\hat{\theta}}^{(s)}}\!\left(\mathbf{X}_i^{(s)} \right)} , \label{eq:rho_hat_cumul}
\end{align}
where we do not use the data $\mathbf{X}_1^{(0)},\ldots, \mathbf{X}_{n_{0}}^{(0)}$ from the initial distribution $Q_{\boldsymbol{\hat{\theta}}^{(0)}} = Q_{\boldsymbol{\eta}}$ because they could potentially increase the variance of the resulting estimator if $Q_{\boldsymbol{\eta}}$ is too different from $Q^*$. The estimator in \eqref{eq:rho_hat_cumul} is an importance sampling estimator of $\rho = \mathbb{E}_{\mu}\!\left[r   \right]$. The potential for the increased variance has been well studied in the importance sampling literature \citep[e.g.,][]{hesterberg1995,owen2000safe}.

\section{Application of the Cross-Entropy Information Criterion}\label{sec:application_CIC}
This section details how the CIC can be useful in practice. 
We first present how the CIC can help choose the number of components, $k$, for a mixture model in conjunction with an expectation-maximization (EM) algorithm that finds the MCE for a given $k$ in the $t^{th}$ iteration, $t=1,\ldots,\tau$. Then, we present the summary of how to use the cumulative version of CIC to iteratively approximate a target distribution. Lastly, we provide a numerical example to illustrate the use of the CIC for approximating an optimal importance sampling distribution. 

\subsection{Mixture model and an EM algorithm}\label{subsec:EM_algo}
To approximate a target distribution, we can consider a parametric mixture model whose parameter dimension $d$ 
determines the model complexity. Parametric mixture models are often used to approximate a posterior density for Bayesian inference \citep{gutmann2016bayesian, Blei2017} and an optimal importance sampling density \citep{botev2013, kurtz2013,Wang2015}. 
The density approximation quality hinges on the number of mixture components, $k$ (or equivalently, the model dimension $d$). Prior studies either assume that $k$ is given \citep{botev2013, kurtz2013} or use a rule of thumb to choose $k$ based on ``some understanding of the structure of the problem at hand'' \citep{Wang2015}.  

 We can use the CIC to select $k$ for any parametric mixture model, considering various parametric component families. For example, exponential families are especially convenient because the MCE can be found by using an expectation-maximization (EM) algorithm \citep{dempster1977maximum}. In this paper, we use the Gaussian mixture model (GMM) for illustration. 
 Appendix B details our version of the EM algorithm to find the MCE. 

\begin{figure}[!h]  
	{\centering 
		\subfigure[Initialization]{ \includegraphics[trim = .361in 0 .5in .25in, clip, width=0.227\linewidth]{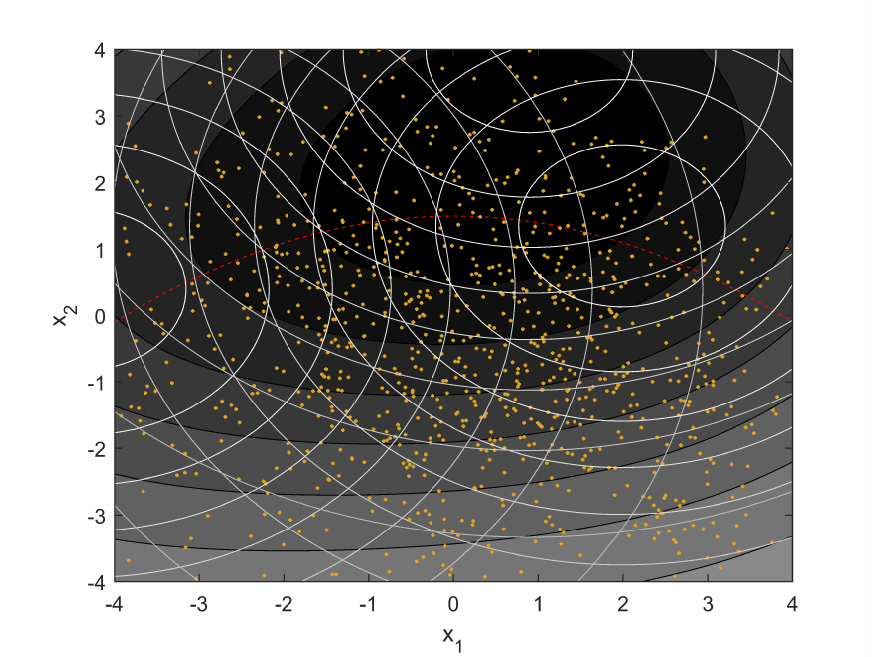} }
		\subfigure[Iteration 1]{ \includegraphics[trim = .361in 0 .5in .25in, clip, width=0.227\linewidth]{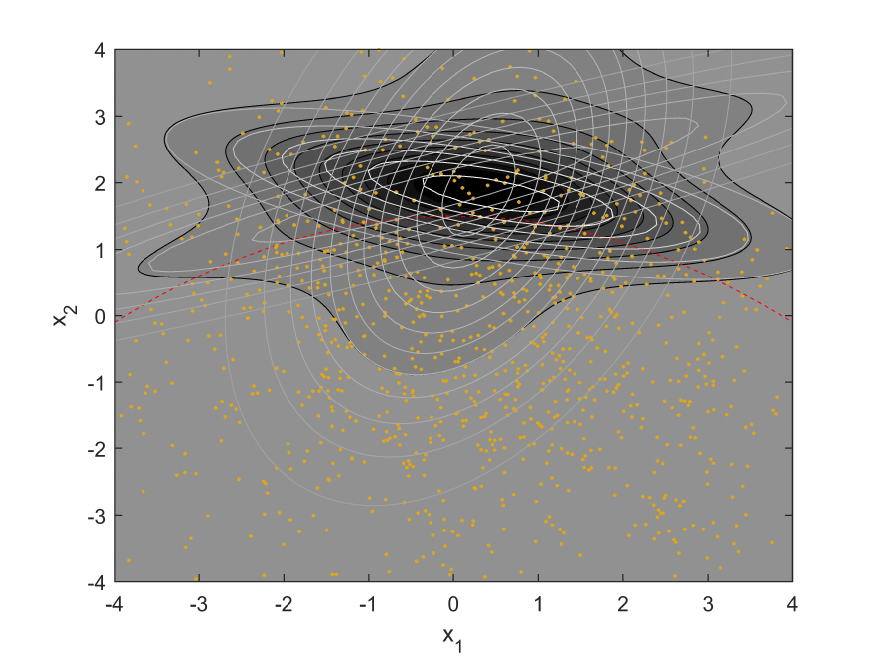} }
		\subfigure[Iteration 2]{ \includegraphics[trim = .361in 0 .5in .25in, clip, width=0.227\linewidth]{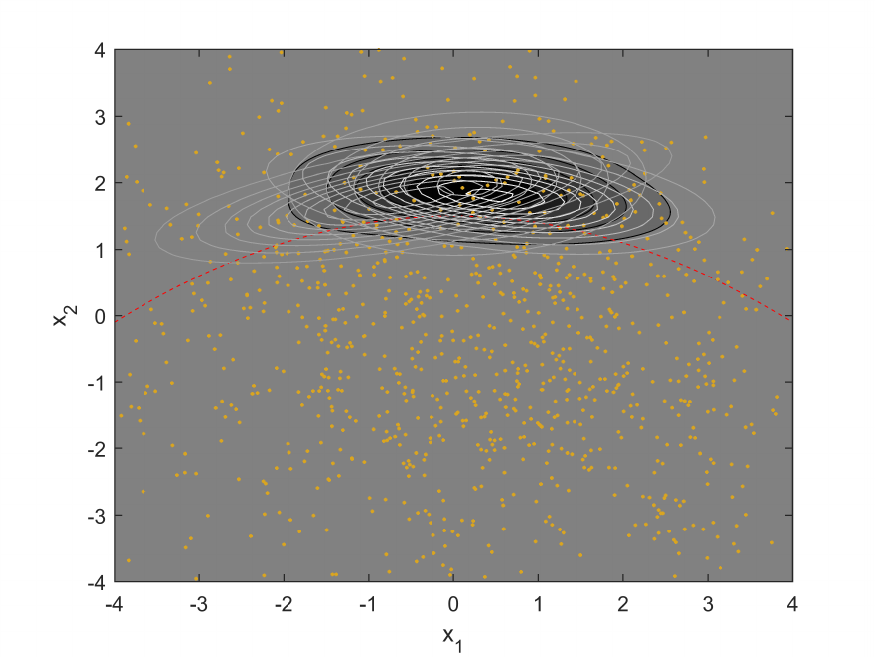}}
		\subfigure[Iteration 3]{ \includegraphics[trim = .361in 0 .5in .25in, clip, width=0.227\linewidth]{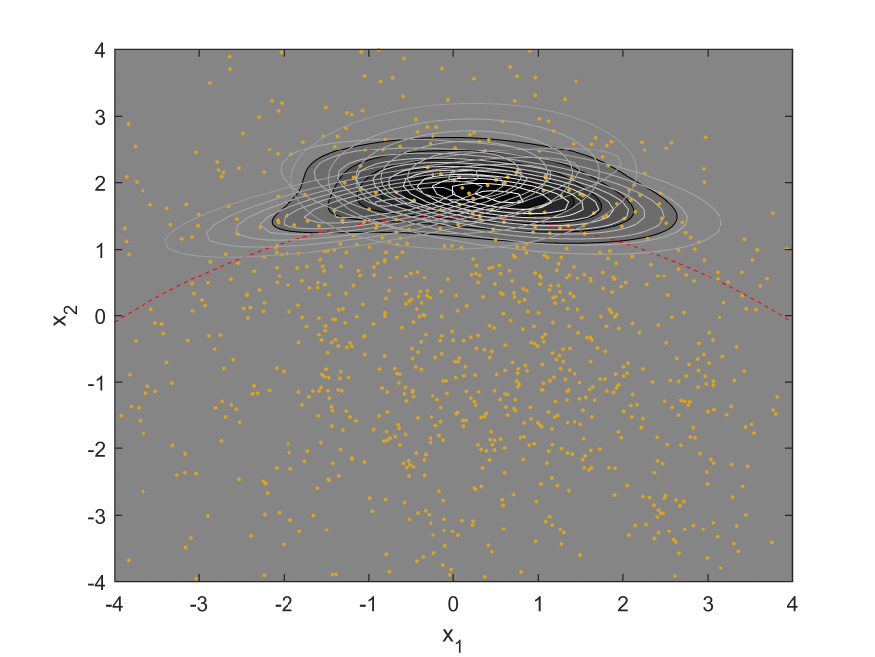}}
	}\vspace{-1em}
	{\centering 
		\subfigure[Iteration 4]{ \includegraphics[trim = .361in 0 .5in .25in, clip, width=0.227\linewidth]{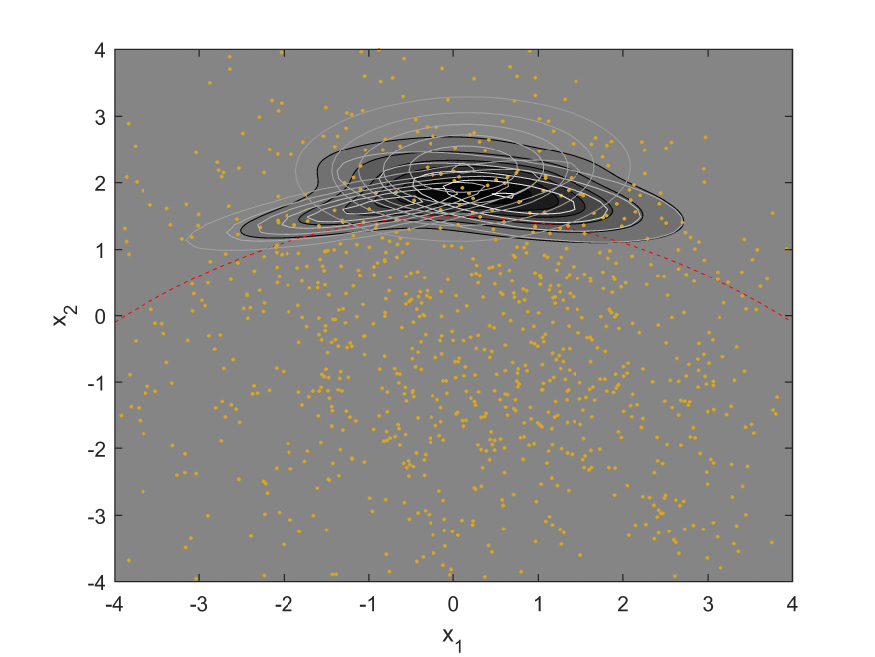} }
		\subfigure[Iteration 5]{ \includegraphics[trim = .361in 0 .5in .25in, clip, width=0.227\linewidth]{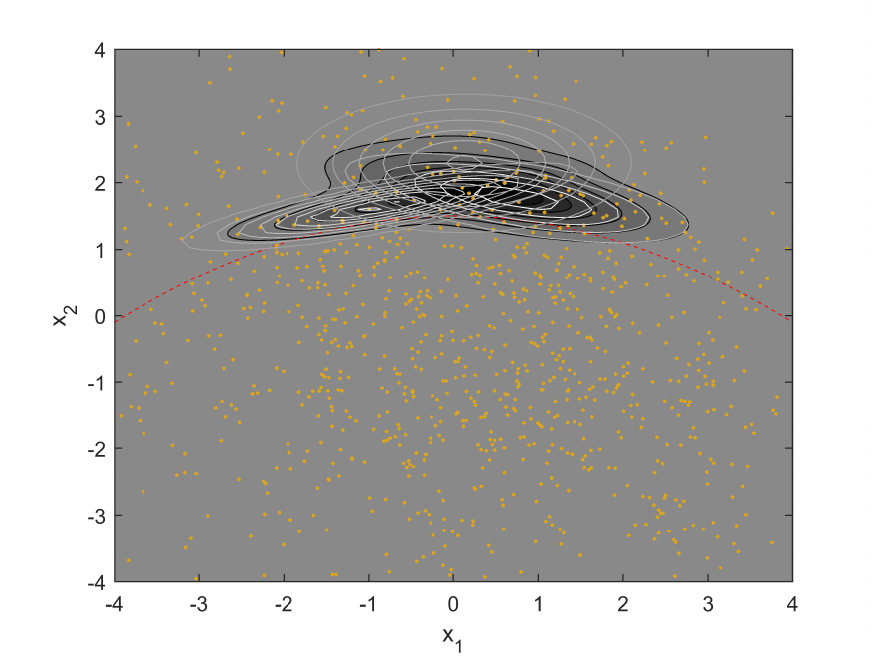} }
		\subfigure[Iteration 6]{ \includegraphics[trim = .361in 0 .5in .25in, clip, width=0.227\linewidth]{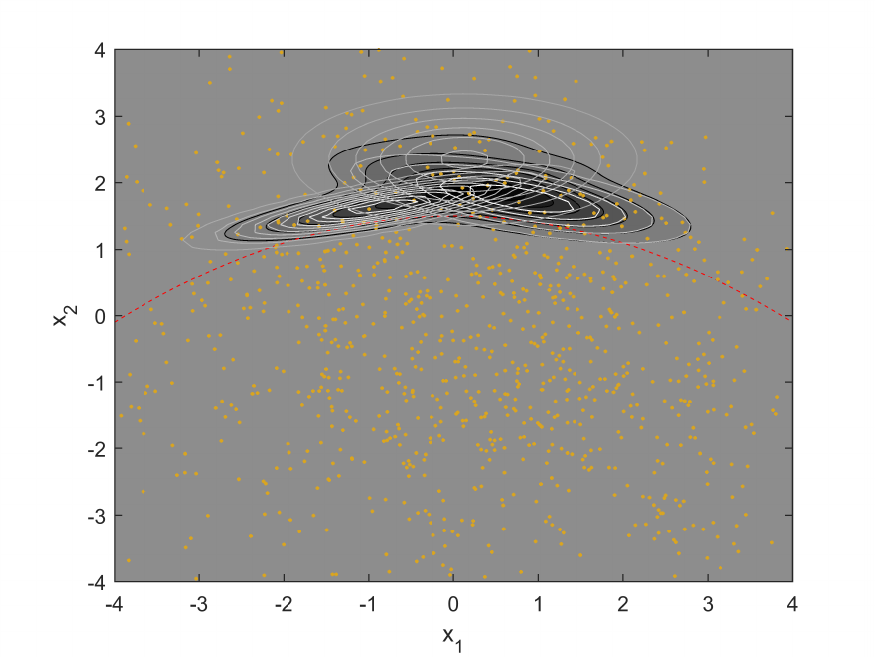}}
		\subfigure[Target density]{ \includegraphics[trim = .361in 0 .5in .25in, clip, width=0.227\linewidth]{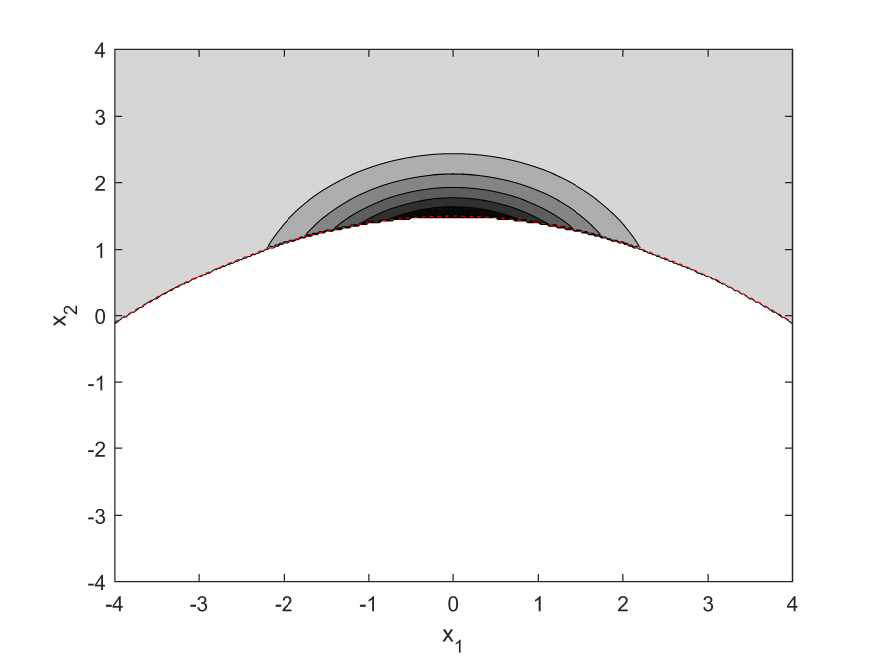}}
	}
	\centering \caption{Illustration of our EM algorithm (for $k=3$ at $t=1$) that updates a randomly initialized density in (a) through Iterations 1-6 in (b)--(g) to approximate the unknown target density in (h). \textit{Yellow dots} in (a)--(g) are the 2-dimensional pilot data $\mathbf{X}_i^{(0)}, i = 1,\ldots, 1000$ sampled from the initial distribution $Q_{\boldsymbol{\hat{\theta}}^{(0)}} = Q_{\boldsymbol{\eta}}$. \textit{Gray-scale filled contour plots} represent the Gaussian mixture density with $k=3$ components in (a)--(g) and the target density in (h). \textit{White contour line plots} in (a)--(g) represent the three component densities of the Gaussian mixture density. \textit{Red dashed line} is the reference line that marks the shape of the target density in (h). The target density is the optimal importance sampling density in the numerical example (with $b=1.5$) in Section~\ref{sec:num_ex}.} \label{fig:EM algo}
\end{figure}

Figure~\ref{fig:EM algo} illustrates our EM algorithm in action for the first outer iteration ($t=1$), where a GMM (gray-scale filled countour plot) with three component densities (white countour lines) is updated over EM iterations to approximate an unknown target density. We can see that in contrast to the conventional EM algorithm that maximizes the likelihood of a model (i.e., goodness-of-fit) to approximate the distribution of observed data, our algorithm uses the data (yellow dots), $\mathbf{X}_1^{(0)}, \ldots, \mathbf{X}_{1000}^{(0)} \sim Q_{\boldsymbol{\eta}}$, to estimate and minimize the cross-entropy from the approximate density to the target density. Out of the 1000 observations (yellow dots) in Figure~\ref{fig:EM algo}(a) (note that the same data are plotted in (b)--(g) as well), only the small portion of them that fall \textit{above} the red dashed line contribute to the cross-entropy estimate (in \eqref{eq:CE_cumulative} because $r(\mathbf{x})$ is zero \textit{below} the red dashed line for the numerical example in Section~\ref{sec:num_ex}). 
See Appendix C for more details of the EM algorithm implemented for the numerical example in Section~\ref{sec:num_ex}. 

\subsection{Summary of the CIC-based distribution approximation procedure}\label{subsec:summary_procedure} 
This subsection summarizes how we can use the CIC to approximate a target distribution in practice. Using the EM algorithm in Section~\ref{subsec:EM_algo} for different $k$'s (or $d$'s) in the $t^{th}$ iteration for $t=1,\ldots,\tau$, we can find the MCE in \eqref{eq:MCE_cumul} and calculate the CIC in \eqref{eq:CIC_cumul}. At the minimum of the CIC, we can then find the best number of components, $k^{*(t)}$ (or the best model dimension $d^{*(t)}$) to use in the $t^{th}$ iteration. 

\begin{figure}[!h]
	{\centering
		\includegraphics[width=4.5in]{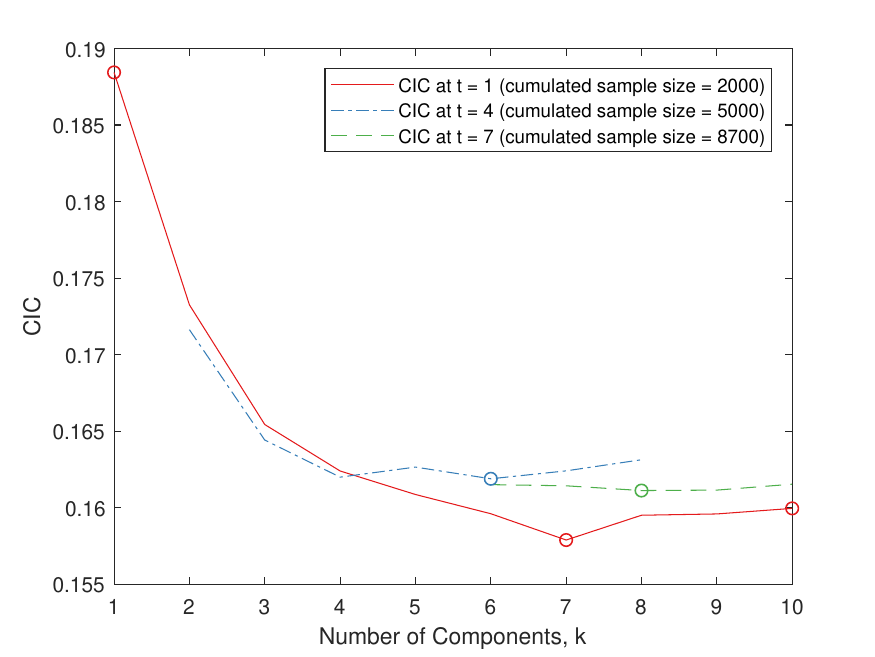} \\
		\centering\caption{Plot of the CIC (cumulative version), $\overline{\textrm{CIC}}^{(t)}(d)$, in \eqref{eq:CIC_cumul} versus the number of components $k$, which determines the model dimension $d$,  of the Gaussian mixture model. As the iteration counter $t$ increases from {\color[rgb]{0.8941176,0.1019608,0.1098039}1 (red solid line)} to {\color[rgb]{0.2156863,0.4941176,0.7215686}4 (black dash-dot line)} to {\color[rgb]{0.3019608,0.6862745,0.2901961}7 (green dashed line)}, the CIC is calculated using a larger sample. The CIC is minimized at $k=7$ for {\color[rgb]{0.8941176,0.1019608,0.1098039} $t=1$}, $k=6$ for {\color[rgb]{0.2156863,0.4941176,0.7215686}$t=4$}, and $k=8$ for {\color[rgb]{0.3019608,0.6862745,0.2901961}$t=7$} in this example. The circles in the plot correspond to the approximate densities shown in Figures~\ref{fig:CIC_illustration2}(a)--(e).}
		\label{fig:CIC_illustration}
	}
\end{figure}

The CIC tends to decrease and then slowly increase as $d$ increases, subject to the randomness of the data. Figure~\ref{fig:CIC_illustration} shows such a pattern, where $k$ is the number of mixture components in the GMM with unconstrained means and covariances. Note that $k$ is proportional to the free parameter dimension $d = (k-1) + k(p + {p(p+1)}/{2})$, with $p$ denoting the dimension of the GMM density support. 

Within the $t^{th}$ iteration, 
we calculate the CIC over $k$ as shown in Figure~\ref{fig:CIC_illustration} for $t = $ {\color[rgb]{0.8941176,0.1019608,0.1098039}1}, {\color[rgb]{0.2156863,0.4941176,0.7215686}4}, {\color[rgb]{0.3019608,0.6862745,0.2901961}7} for the numerical example (with $b=1.5$) in Section~\ref{sec:num_ex}. As the iteration counter $t$ increases, $\overline{\textrm{CIC}}^{(t)}(d)$ in \eqref{eq:CIC_cumul} uses a larger sample that accumulated data over iterations to more accurately estimate the cross-entropy. 

\begin{figure}[!h]
	\centering  
	{\centering 
		\subfigure[{GMM ($k=1$) at \color[rgb]{0.8941176,0.1019608,0.1098039} $t=1$}]{ \includegraphics[trim = .361in 0 .5in .25in, clip, width=0.306\linewidth]{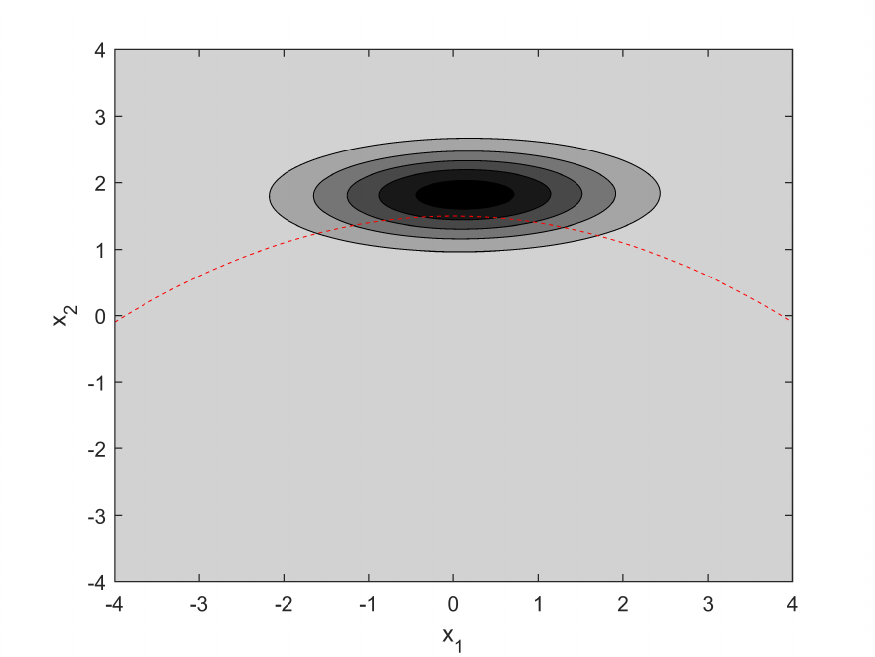} }
		\subfigure[GMM ($k=7$) at {\color[rgb]{0.8941176,0.1019608,0.1098039} $t=1$}]{ \includegraphics[trim = .361in 0 .5in .25in, clip, width=0.306\linewidth]{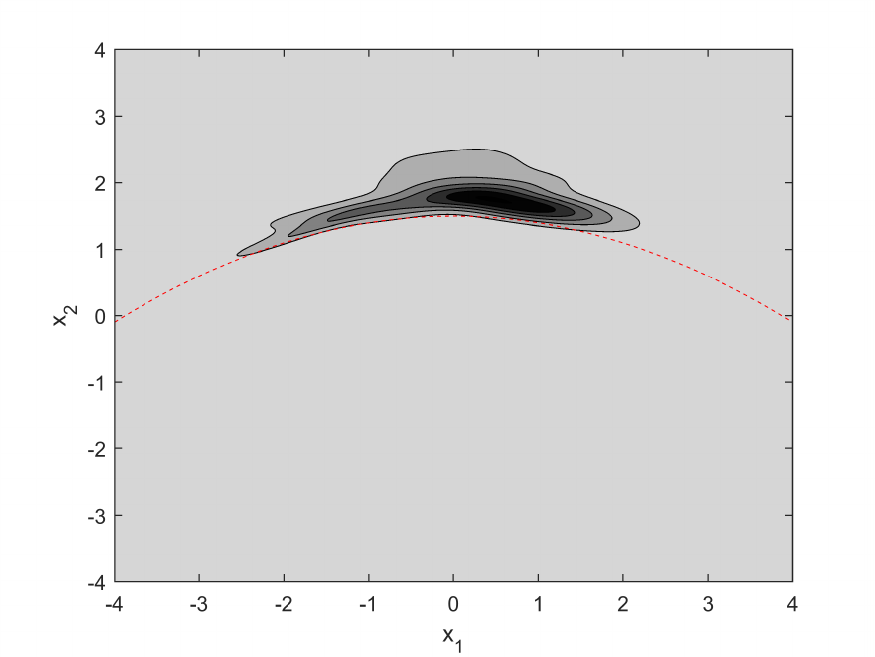} }
		\subfigure[GMM ($k=10$) at {\color[rgb]{0.8941176,0.1019608,0.1098039} $t=1$}]{ \includegraphics[trim = .361in 0 .5in .25in, clip, width=0.306\linewidth]{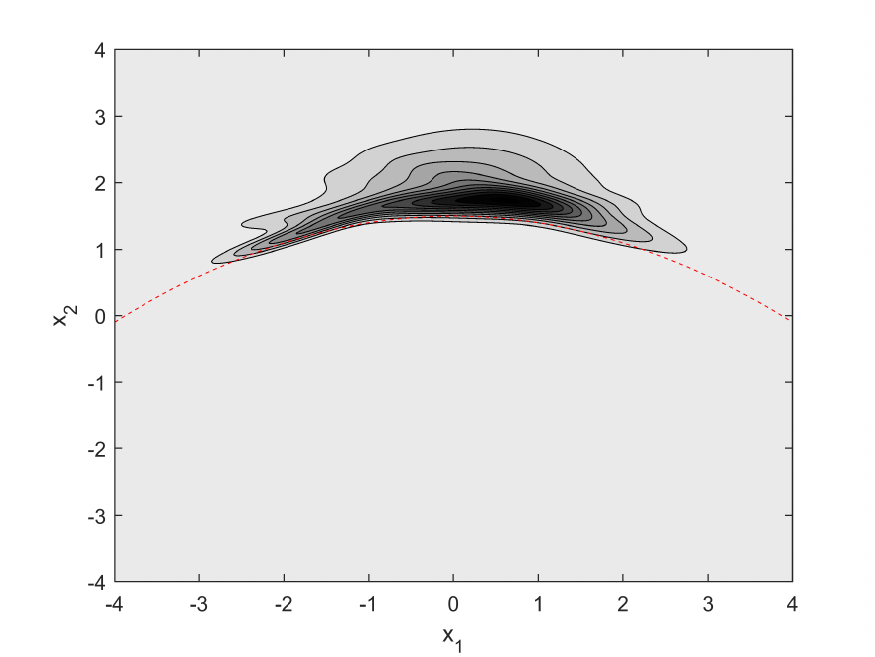} }
	}\vspace{-1em}
	{\centering 
		\subfigure[GMM ($k=6$) at {\color[rgb]{0.2156863,0.4941176,0.7215686} $t=4$}]{ \includegraphics[trim = .361in 0 .5in .25in, clip, width=0.306\linewidth]{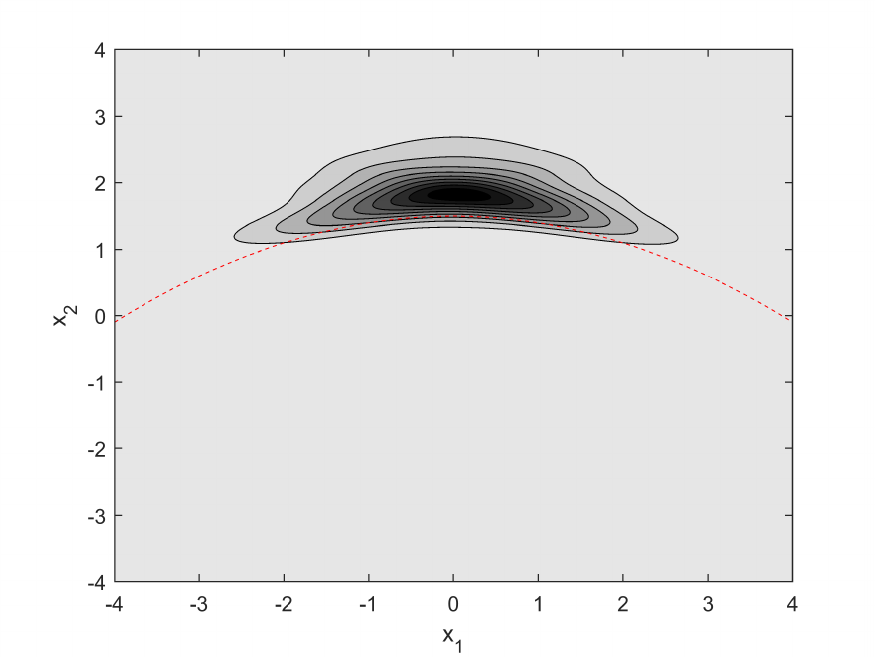} }
		\subfigure[GMM ($k=8$) at {\color[rgb]{0.3019608,0.6862745,0.2901961} $t=7$}]{ \includegraphics[trim = .361in 0 .5in .25in, clip, width=0.306\linewidth]{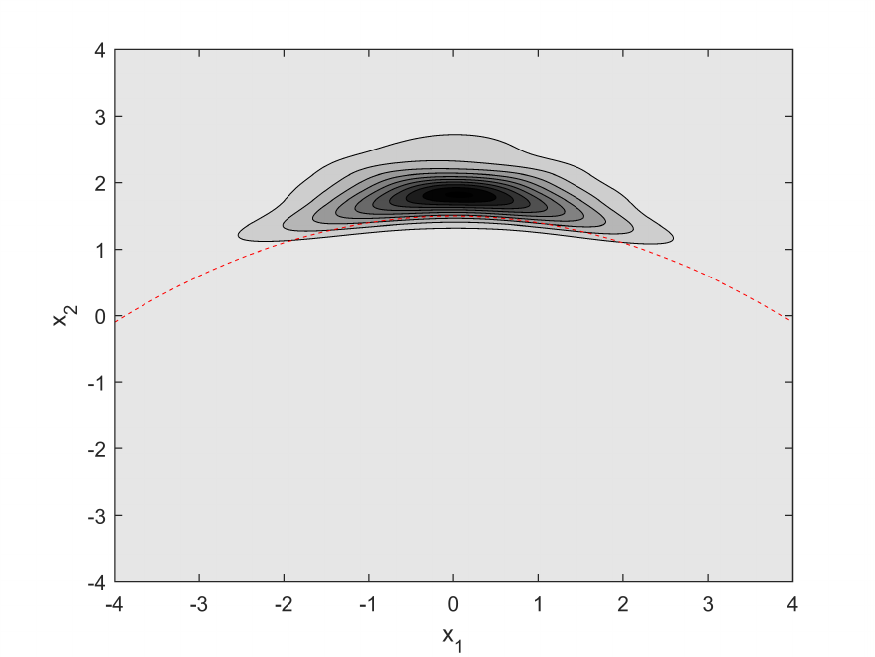} }
		\subfigure[Target density]{ \includegraphics[trim = .361in 0 .5in .25in, clip, width=0.306\linewidth]{CIC-IS_optimal_density_bb15.pdf} }
	} 
	\centering \caption{Gaussian mixture models (GMMs) with $k$ components in (a)--(e) correspond to the circles in Figure~\ref{fig:CIC_illustration} and are compared with the target density in (f), which is the optimal importance sampling density in the numerical example (with $b=1.5$) in Section~\ref{sec:num_ex}.} \label{fig:CIC_illustration2}
\end{figure}

At {\color[rgb]{0.8941176,0.1019608,0.1098039} $t=1$}, the GMM with $k=7$ in Figure~\ref{fig:CIC_illustration2}(b) achieves the minimum CIC (as shown in Figure~\ref{fig:CIC_illustration}), while $k=1$ and $k=10$ result in seemingly over-simplified and over-complicated densities in Figure~\ref{fig:CIC_illustration2}(a) and Figure~\ref{fig:CIC_illustration2}(c), respectively, for the given sample size, 1000 (note that the effective sample size is much smaller because only a small portion of the data fall above the red dashed line as explained earlier with Figure~\ref{fig:EM algo}). 
The right choice of $k$ (neither too small nor too large) for the given sample size at the current iteration helps subsequent iterations by preventing sampling from an overly simplified/complicated distribution that could misguide the later iterations. 
Thus, it is beneficial to refine the approximate distribution \textit{proportionally} (neither too much nor too little) for the given data size. 
Over iterations, sampled data (yellow dots in Figure~\ref{fig:EM algo}) should increasingly cover the entire support of the target density. 
The CIC-minimizing densities at {\color[rgb]{0.2156863,0.4941176,0.7215686}$t=4$} in Figure~\ref{fig:CIC_illustration2}(d) and {\color[rgb]{0.3019608,0.6862745,0.2901961}$t=7$} in Figure~\ref{fig:CIC_illustration2}(e) capture the overall shape of the target density in Figure~\ref{fig:CIC_illustration2}(f). 



\begin{figure}[!h]
	\fbox{\parbox{5.9in}{
			\begin{center}
				{\sc CIC-Based Approximation of the Target Distribution $Q^*$}
			\end{center}
			Inputs: iteration counter $t=1$, the number of iterations $\tau$, the sample size per iteration $n_{t}, t=1,\ldots,\tau$, the initial parameter dimension $d^{(0)}$, and the initial parameter $\boldsymbol{\hat{\theta}}^{(0)} = \boldsymbol{\eta} \in \boldsymbol{\Theta}_{d^{(0)}}$.
			\begin{center}
				\begin{enum}
					\item[1.] Sample $\mathbf{X}_{1}^{(t-1)},\ldots, \mathbf{X}_{n_{t-1}}^{(t-1)} \sim Q_{\boldsymbol{\hat{\theta}}^{(t-1)}}$.
					\item[2.] Find the best model dimension $d^{*(t)}:= \underset{d\ge 1}{\arg\!\min}\,\overline{\textrm{CIC}}^{(t)}(d)$ to use,  
					where 							
					$\overline{\textrm{CIC}}^{(t)}(d) := \bar{\mathcal{C}}^{(t-1)}\!\left(\boldsymbol{\hat{\theta}}^{(t)}\right) + \hat{\rho}\frac{d}{\sum_{s = 0}^{t-1}n_{s}}$ is in \eqref{eq:CIC_cumul} with the MCE 
					$\boldsymbol{\hat{\theta}}^{(t)} := \underset{\boldsymbol{\theta}\in \boldsymbol{\Theta}_d}{\arg\!\min}\,\bar{\mathcal{C}}^{(t-1)}(\boldsymbol{\theta})$ in \eqref{eq:MCE_cumul}. 
					$\hat{\rho}$ is a consistent estimator of $\rho$, such as the estimators in \eqref{eq:rho_hat_cumul0} and \eqref{eq:rho_hat_cumul}. 	
					\item[3.] If $t = \tau$, output the approximate distribution  $Q_{\boldsymbol{\hat{\theta}}^{(\tau)}}$.  Otherwise, increment $t$ by 1 and go to Step 1.  
				\end{enum}
			\end{center}
	}}
	\caption{CIC-based approximation of the target distribution $Q^*$}
	\label{fig:dist_approx}
\end{figure}
Figure~\ref{fig:dist_approx} summarizes the CIC-based distribution approximation procedure. 
Note that in addition to approximating the target distribution $Q^*$, if we want to estimate a quantity of interest such as $\rho$, we can sample $\mathbf{X}_{1}^{(\tau)},\ldots, \mathbf{X}_{n_{\tau}}^{(\tau)} \sim Q_{\boldsymbol{\hat{\theta}}^{(\tau)}}$ and use the estimator such as $\hat{\rho}^{(\tau)}$ in \eqref{eq:rho_hat_cumul}. 
The following numerical example in Section~\ref{sec:num_ex} uses this additional step for importance sampling. 

\subsection{Numerical example}\label{sec:num_ex}



This subsection empirically demonstrates how the CIC can be used to approximate the optimal distribution of importance sampling for estimating $\rho$. We expect that the better the approximation is, the smaller the standard error of the estimator $\hat{\rho}^{(\tau)}$ in \eqref{eq:rho_hat_cumul} will be. We call this importance sampling method the CIC-IS. 


As benchmarks, we consider two methods to estimate $\rho$. First, the crude Monte Carlo (CMC) method is the most straightforward and most widely used Monte Carlo method. Second, we implement the importance sampling method in \citet{kurtz2013}, which represents a state-of-the-art, cross-entropy method using a mixture model. Their method, which is called cross-entropy-based adaptive importance sampling using Gaussian mixture (CE-AIS-GM) uses the GMM with a \textit{pre-specified} value for the number of mixture components, $k$. The GMM parameters are updated \textit{once} within each iteration (as in Figure~\ref{fig:EM algo}(b)) instead of using an EM algorithm. 

We note that this empirical comparison is to \textit{illustrate} how the CIC can improve importance sampling (by improving the approximation of the optimal density), \textit{not} to comprehensively test the empirical performance of the CIC-IS itself. 
Interested readers are referred to the work of \cite{cao2019cross}, which applies the CIC-IS to stochastic simulation models and studies its empirical performance using extensive numerical experiments and a case study on reliability evaluation of a wind turbine.

We use a classical example in the structural safety literature, which is also used in \cite{kurtz2013}. With $\mathbf{X}$ following the bivariate Gaussian density $\phi(\mathbf{x})$ with zero mean and identity covariance matrix, a system of interest fails when $\mathbf{X}$ falls on the region $\{\mathbf{x} \in \mathbb{R}^2 : g(\mathbf{x}) \leq 0\}$, where 
$g(\mathbf{x})  = b - x_2 - \kappa \left(x_1 - e \right)^2$.
We vary the parameter $b = 1.5$, $2.0$, $2.5$ to test three different failure thresholds. We fix the other two parameters, $\kappa = 0.1$ and $e = 0$, to maintain the shape of the failure boundary $\{\mathbf{x} \in \mathbb{R}^2 : g(\mathbf{x}) = 0\}$ (the red dashed line in Figure~\ref{fig:CIC_illustration2}). Note that $g(\mathbf{x})$ \textit{represents} a computationally expensive function to evaluate, such as a finite element model in structural engineering.  

The quantity of interest is the probability of the failure event, $\rho = \mathbb{E}_{\mu}\!\left[r   \right]$, where $r(\mathbf{x}) = \phi(\mathbf{x}) \mathbb{I}(g(\mathbf{x}) \leq 0).$ Here, $\mathbb{I}(\cdot)$ denotes the indicator function. The CMC estimator of $\rho$ is 
\begin{align}
\hat{\rho}_{\textrm{CMC}} = \frac{1}{n_{\textrm{CMC}}}\sum_{i=1}^{n_{\textrm{CMC}}}  \mathbb{I}(g(\mathbf{X}_i) \leq 0)  , \label{eq:rho_cmc} 
\end{align}
where $\mathbf{X}_i, i = 1,\ldots, n_{\textrm{CMC}}$, are sampled from the density $\phi(\mathbf{x})$. 
For both importance sampling methods, we use the same sample size in \cite{kurtz2013}, namely, the total of 8700 replications: $n_t = 1000$ for $t = 0, \ldots, 6$ and $n_\tau = 1700$ for $\tau = 7$. Because the standard error of the CMC estimator can be analytically calculated as $\rho (1-\rho)/n_{\textrm{CMC}}$, we estimate the standard error instead of running CMC simulations. 


We set the CE-AIS-GM to use $k=30$ and to estimate $\rho$ based only on the last ($\tau^{(th)}$) iteration data as in \citet{kurtz2013}. The CIC-IS adaptively chooses $k$ within the algorithm described in Figure~\ref{fig:dist_approx} and uses the data from multiple iterations ($t = 2, \ldots, \tau$) to estimate $\rho$ by $\hat{\rho}^{(\tau)}$ in \eqref{eq:rho_hat_cumul}, as described in Section~\ref{subsec:summary_procedure}. Because the CIC helps find a distribution fairly close to the optimal distribution throughout all the iterations, the CIC-IS can use the accumulated data to estimate $\rho$. 

Table~\ref{tab:ex1} shows the estimation results based on 500 experiment repetitions. As the parameter $b$ increases, the estimand $\rho$ decreases. 
Regardless of $b$, the CIC-IS obtains at least 50\% smaller standard errors than the CE-AIS-GM. 
The smaller standard errors translate into larger computational savings for estimating $\rho$ at a desired accuracy. Specifically, to compare both importance sampling methods with CMC, we analytically calculate `CMC Ratio', which is the number of replications used in each row's method (that is, 8700) divided by the number of replications necessary for the CMC estimator in \eqref{eq:rho_cmc} (that is, $n_{\textrm{CMC}}$) to achieve the same standard error in the row. Although CE-AIS-GM saves significantly compared to CMC, CIC-IS saves even more by 4 to 6 times.
The good performance of the CIC-IS can be attributed to a) closeness of the approximate densities $q_{\boldsymbol{\hat{\theta}}^{(t)}}$, $t=1,\ldots,\tau$ to the optimal density $q^*$ as illustrated in Figure~\ref{fig:CIC_illustration2}, and b) use of the data from multiple iterations ($t = 2, \ldots, \tau$) to estimate $\rho$, thanks to good approximate densities throughout the iterations.


\begin{table}[h]
	\centering
	\caption{Comparison between CE-AIS-GM and CIC-IS}
	\label{tab:ex1}
	\begin{tabular}{ccccc}
		\hline
		$b$ & Method    & Mean     & Standard Error & CMC Ratio \\ \hline
		1.5 & CE-AIS-GM & 0.082902 & 0.001145       & 15.00\%     \\
		& CIC-IS      & 0.082911 & 0.000506       & 2.93\%      \\
		2.0 & CE-AIS-GM & 0.030174 & 0.000526       & 8.23\%      \\
		& CIC-IS      & 0.030173 & 0.000213       & 1.35\%      \\
		2.5 & CE-AIS-GM & 0.008908 & 0.000211       & 4.39\%      \\
		& CIC-IS      & 0.008910 & 0.000099       & 0.97\%      \\ \hline
	\end{tabular}
	\\
	\raggedright
	{\footnotesize Note: `Mean' and `Standard Error' are the sample mean and standard error of the estimates, respectively. The `CMC Ratio' is $n_{\textrm{Total}} / n_{\textrm{CMC}}$, where $n_{\textrm{Total}} =8700$ and $n_{\textrm{CMC}}= \bar{\rho}(1-\bar{\rho})/(S.E.)^2$. $\bar{\rho}$ is the CIC-IS's sample mean and $S.E.$ is the standard error of the method in the row. The smaller the CMC Ratio, the larger the computational saving of the method over CMC.
	}
\end{table}

\section{Conclusion}\label{sec:conclusion}
This paper proposed a novel information criterion, called the cross-entropy information criterion (CIC), to find a parametric density that has the asymptotically minimum cross-entropy to a target density to approximate. The CIC is the sum of two terms: an estimator of the cross-entropy (up to a multiplicative constant) from the parametric density to the target density, and a model complexity penalty term that is proportional to the free parameter dimension of the parametric density. Under certain regularity conditions, we proved that the penalty term corrects the asymptotic bias of the first term in estimating the true cross-entropy. Empirically, we demonstrated that minimizing the CIC leads to a density that well approximates an optimal importance sampling density. 
The CIC allowed us to develop a principled algorithm to automatically approximate an unknown density that can be evaluated up to a normalizing constant at a limited number of points. This opens up the possibility to create an off-the-shelf software package for importance sampling and other similar applications. 

\vskip 14pt
\noindent {\large\bf Supplementary Materials}

The online supplement includes Appendix A (Assumptions and Proofs), Appendix B (EM Algorithm for Minimizing the Cross-Entropy), and Appendix C (Implementation Details of the Numerical Example). 
\par
\vskip 14pt
\noindent {\large\bf Acknowledgements}

This work was supported by the National Science Foundation (NSF grant DMS-1952781). 
\par


\markboth{\hfill{\footnotesize\rm YOUNGJUN CHOE, YEN-CHI CHEN, and \textcolor{black}{NICK TERRY}} \hfill}
{\hfill {\footnotesize\rm CROSS-ENTROPY INFORMATION CRITERION} \hfill}

\bibhang=1.7pc
\bibsep=2pt
\fontsize{9}{14pt plus.8pt minus .6pt}\selectfont
\renewcommand\bibname{\large \bf References}
\expandafter\ifx\csname
natexlab\endcsname\relax\def\natexlab#1{#1}\fi
\expandafter\ifx\csname url\endcsname\relax
  \def\url#1{\texttt{#1}}\fi
\expandafter\ifx\csname urlprefix\endcsname\relax\def\urlprefix{URL}\fi

\lhead[\footnotesize\thepage\fancyplain{}\leftmark]{}\rhead[]{\fancyplain{}\rightmark\footnotesize{} }
%
%
%
%
%
\bibliographystyle{plainnat}
\bibliography{CIC_bib}

\begin{thebibliography}{33}
\providecommand{\natexlab}[1]{#1}
\providecommand{\url}[1]{\texttt{#1}}
\expandafter\ifx\csname urlstyle\endcsname\relax
  \providecommand{\doi}[1]{doi: #1}\else
  \providecommand{\doi}{doi: \begingroup \urlstyle{rm}\Url}\fi

\bibitem[Akaike(1974)]{akaike1974}
Hirotugu Akaike.
\newblock A new look at the statistical model identification.
\newblock \emph{IEEE Transactions on Automatic Control}, 19\penalty0
  (6):\penalty0 716--723, 1974.

\bibitem[Beaumont(2010)]{Beaumont2010}
Mark~A. Beaumont.
\newblock Approximate {Bayesian} computation in evolution and ecology.
\newblock \emph{Annual Review of Ecology, Evolution, and Systematics},
  41:\penalty0 379--406, 2010.

\bibitem[Bhansali(1986)]{bhansali1986derivation}
Rajendra~J. Bhansali.
\newblock A derivation of the information criteria for selecting autoregressive
  models.
\newblock \emph{Advances in Applied Probability}, 18\penalty0 (2):\penalty0
  360--387, 1986.

\bibitem[Bhansali and Papangelou(1991)]{bhansali1991convergence}
Rajendra~J. Bhansali and Fredos Papangelou.
\newblock Convergence of moments of least squares estimators for the
  coefficients of an autoregressive process of unknown order.
\newblock \emph{The Annals of Statistics}, 19\penalty0 (3):\penalty0
  1155--1162, 1991.

\bibitem[Blei et~al.(2017)Blei, Kucukelbir, and McAuliffe]{Blei2017}
David~M. Blei, Alp Kucukelbir, and Jon~D. McAuliffe.
\newblock Variational inference: A review for statisticians.
\newblock \emph{Journal of the American Statistical Association}, 112\penalty0
  (518):\penalty0 859--877, 2017.

\bibitem[Botev et~al.(2013)Botev, Kroese, Rubinstein, and L'Ecuyer]{botev2013}
Zdravko~I. Botev, Dirk~P. Kroese, Reuven~Y. Rubinstein, and Pierre L'Ecuyer.
\newblock The cross-entropy method for optimization.
\newblock \emph{Machine Learning: Theory and Applications, V. Govindaraju and
  C. R. Rao, Eds, Chennai: Elsevier}, 31:\penalty0 35--59, 2013.

\bibitem[Cao and Choe(2019)]{cao2019cross}
Quoc~Dung Cao and Youngjun Choe.
\newblock Cross-entropy based importance sampling for stochastic simulation
  models.
\newblock \emph{Reliability Engineering \& System Safety}, 191:\penalty0
  106526, 2019.

\bibitem[Chen and Choe(2019)]{chen2019importance}
Yen-Chi Chen and Youngjun Choe.
\newblock Importance sampling and its optimality for stochastic simulation
  models.
\newblock \emph{Electronic Journal of Statistics}, 13\penalty0 (2):\penalty0
  3386--3423, 2019.

\bibitem[Choe et~al.(2015)Choe, Byon, and Chen]{Choe2015}
Youngjun Choe, Eunshin Byon, and Nan Chen.
\newblock Importance sampling for reliability evaluation with stochastic
  simulation models.
\newblock \emph{Technometrics}, 57\penalty0 (3):\penalty0 351--361, 2015.

\bibitem[Claeskens and Consentino(2008)]{Claeskens2008}
Gerda Claeskens and Fabrizio Consentino.
\newblock Variable selection with incomplete covariate data.
\newblock \emph{Biometrics}, 64\penalty0 (4):\penalty0 1062--1069, 2008.
\newblock \doi{10.1111/j.1541-0420.2008.01003.x}.

\bibitem[Dempster et~al.(1977)Dempster, Laird, and Rubin]{dempster1977maximum}
Arthur~P. Dempster, Nan~M. Laird, and Donald~B. Rubin.
\newblock Maximum likelihood from incomplete data via the {EM} algorithm.
\newblock \emph{Journal of the Royal Statistical Society. Series B
  (Methodological)}, pages 1--38, 1977.

\bibitem[Donohue et~al.(2011)Donohue, Overholser, Xu, and
  Vaida]{donohue2011conditional}
Michael~C. Donohue, Rosanna Overholser, Ronghui Xu, and Florin Vaida.
\newblock Conditional {Akaike} information under generalized linear and
  proportional hazards mixed models.
\newblock \emph{Biometrika}, 98\penalty0 (3):\penalty0 685--700, 2011.

\bibitem[Figueiredo and Jain(2002)]{figueiredo2002}
Mario A.~T. Figueiredo and Anil~K. Jain.
\newblock Unsupervised learning of finite mixture models.
\newblock \emph{IEEE Transactions on Pattern Analysis and Machine
  Intelligence}, 24\penalty0 (3):\penalty0 381--396, 2002.

\bibitem[Findley and Wei(2002)]{findley2002aic}
David~F. Findley and Ching-Zong Wei.
\newblock {AIC}, overfitting principles, and the boundedness of moments of
  inverse matrices for vector autotregressions and related models.
\newblock \emph{Journal of Multivariate Analysis}, 83\penalty0 (2):\penalty0
  415--450, 2002.

\bibitem[Gutmann and Corander(2016)]{gutmann2016bayesian}
Michael~U. Gutmann and Jukka Corander.
\newblock Bayesian optimization for likelihood-free inference of
  simulator-based statistical models.
\newblock \emph{Journal of Machine Learning Research}, 17\penalty0
  (125):\penalty0 1--47, 2016.

\bibitem[Hastings(1970)]{hastings1970monte}
Wilfred~K. Hastings.
\newblock {Monte Carlo} sampling methods using {Markov} chains and their
  applications.
\newblock \emph{Biometrika}, 57\penalty0 (1):\penalty0 97--109, 1970.

\bibitem[Hesterberg(1995)]{hesterberg1995}
Tim Hesterberg.
\newblock Weighted average importance sampling and defensive mixture
  distributions.
\newblock \emph{Technometrics}, 37\penalty0 (2):\penalty0 185--194, 1995.

\bibitem[Huber(1964)]{huber1964robust}
Peter~J. Huber.
\newblock Robust estimation of a location parameter.
\newblock \emph{The Annals of Mathematical Statistics}, 35\penalty0
  (1):\penalty0 73--101, 1964.

\bibitem[Kahn and Marshall(1953)]{kahn1953}
Herman Kahn and Andy~W. Marshall.
\newblock Methods of reducing sample size in {Monte Carlo} computations.
\newblock \emph{Journal of the Operations Research Society of America},
  1\penalty0 (5):\penalty0 263--278, 1953.

\bibitem[Knuth et~al.(2015)Knuth, Habeck, Malakar, Mubeen, and
  Placek]{knuth2015bayesian}
Kevin~H. Knuth, Michael Habeck, Nabin~K. Malakar, Asim~M. Mubeen, and Ben
  Placek.
\newblock Bayesian evidence and model selection.
\newblock \emph{Digital Signal Processing}, 47:\penalty0 50--67, 2015.

\bibitem[Kullback and Leibler(1951)]{kullback1951information}
Solomon Kullback and Richard~A. Leibler.
\newblock On information and sufficiency.
\newblock \emph{The Annals of Mathematical Statistics}, 22\penalty0
  (1):\penalty0 79--86, 1951.

\bibitem[Kurtz and Song(2013)]{kurtz2013}
Nolan Kurtz and Junho Song.
\newblock Cross-entropy-based adaptive importance sampling using gaussian
  mixture.
\newblock \emph{Structural Safety}, 42:\penalty0 35--44, 2013.

\bibitem[Metropolis et~al.(1953)Metropolis, Rosenbluth, Rosenbluth, Teller, and
  Teller]{metropolis1953equation}
Nicholas Metropolis, Arianna~W. Rosenbluth, Marshall~N. Rosenbluth, Augusta~H.
  Teller, and Edward Teller.
\newblock Equation of state calculations by fast computing machines.
\newblock \emph{The Journal of Chemical Physics}, 21\penalty0 (6):\penalty0
  1087--1092, 1953.

\bibitem[Neddermeyer(2009)]{neddermeyer2009computationally}
Jan~C. Neddermeyer.
\newblock Computationally efficient nonparametric importance sampling.
\newblock \emph{Journal of the American Statistical Association}, 104\penalty0
  (486):\penalty0 788--802, 2009.

\bibitem[Owen and Zhou(2000)]{owen2000safe}
Art Owen and Yi~Zhou.
\newblock Safe and effective importance sampling.
\newblock \emph{Journal of the American Statistical Association}, 95\penalty0
  (449):\penalty0 135--143, 2000.

\bibitem[Pratola and Chkrebtii(2018)]{pratola2018bayesian}
Mathew~T Pratola and Oksana Chkrebtii.
\newblock Bayesian calibration of multistate stochastic simulators.
\newblock \emph{Statistica Sinica}, 28:\penalty0 693--719, 2018.

\bibitem[Rubin(2005)]{rubin2005causal}
Donald~B Rubin.
\newblock Causal inference using potential outcomes: Design, modeling,
  decisions.
\newblock \emph{Journal of the American Statistical Association}, 100\penalty0
  (469):\penalty0 322--331, 2005.

\bibitem[Rubinstein(1999)]{rubinstein1999}
Reuven~Y. Rubinstein.
\newblock The cross-entropy method for combinatorial and continuous
  optimization.
\newblock \emph{Methodology and Computing in Applied Probability}, 1\penalty0
  (2):\penalty0 127--190, 1999.

\bibitem[Rubinstein and Shapiro(1993)]{rubinstein1993}
Reuven~Y. Rubinstein and Alexander Shapiro.
\newblock \emph{Discrete event systems: Sensitivity analysis and stochastic
  optimization by the score function method}.
\newblock Chichester: John Wiley \& Sons Ltd, 1993.

\bibitem[Sunn{\aa}ker et~al.(2013)Sunn{\aa}ker, Busetto, Numminen, Corander,
  Foll, and Dessimoz]{sunnaaker2013approximate}
Mikael Sunn{\aa}ker, Alberto~Giovanni Busetto, Elina Numminen, Jukka Corander,
  Matthieu Foll, and Christophe Dessimoz.
\newblock Approximate {Bayesian} computation.
\newblock \emph{PLOS Computational Biology}, 9\penalty0 (1):\penalty0 1--10, 01
  2013.

\bibitem[Van~der Vaart(1998)]{van1998asymptotic}
Aad~W. Van~der Vaart.
\newblock \emph{Asymptotic statistics}.
\newblock New York: Cambridge University Press, 1998.

\bibitem[Wang and Zhou(2015)]{Wang2015}
Hui Wang and Xiang Zhou.
\newblock A cross-entropy scheme for mixtures.
\newblock \emph{ACM Transactions on Modeling and Computer Simulation},
  25\penalty0 (1):\penalty0 6:1--6:20, 2015.

\bibitem[Zhang(1996)]{zhang1996nonparametric}
Ping Zhang.
\newblock Nonparametric importance sampling.
\newblock \emph{Journal of the American Statistical Association}, 91\penalty0
  (435):\penalty0 1245--1253, 1996.

\end{thebibliography}

\vskip .65cm
\noindent
Youngjun Choe\\ 
Department of Industrial and Systems Engineering\\
University of Washington\\
Seattle, WA 98195-2650, USA
\vskip 2pt
\noindent
E-mail: ychoe@uw.edu 
\vskip 15pt

\noindent
Yen-Chi Chen\\
Department of Statistics\\
University of Washington\\
Seattle, WA 98195-4322, USA
\vskip 2pt
\noindent
E-mail: yenchic@uw.edu 
\vskip 15pt

\noindent
{\color{black}Nick Terry\\ 
Department of Industrial and Systems Engineering\\
University of Washington\\
Seattle, WA 98195-2650, USA
\vskip 2pt
\noindent
E-mail: pnterry@uw.edu}

\newpage
\markboth{\hfill{\footnotesize\rm YOUNGJUN CHOE, YEN-CHI CHEN, and \textcolor{black}{NICK TERRY}} \hfill}
{\hfill {\rm SUPPLEMENT}}
\fontsize{12}{14pt plus.8pt minus .6pt}\selectfont
\renewcommand{\theequation}{S.\arabic{equation}}
\section*{Appendix A: Assumptions and Proofs}

We recall that the functions, $\mathcal{C}(\boldsymbol{\theta}) := - \mathbb{E}_{\mu}\!\left[ r \log{q_{\boldsymbol{\theta}}} \right]$ and $h\!\left(\mathbf{X}, \boldsymbol{\eta}, \boldsymbol{\theta}\right) := \frac{r(\mathbf{X})}{q_{\boldsymbol{\eta}}\!\left(\mathbf{X} \right)}   \log{q_{\boldsymbol{\theta}}\!\left(\mathbf{X} \right)}$, are defined in \eqref{eq:C_theta} and \eqref{eq:func_h}, respectively. Below, $\nabla_{\boldsymbol{\theta}}$ and $\nabla_{\boldsymbol{\theta}}^2$ denote the gradient and the Hessian with respect to $\boldsymbol{\theta}$, respectively. For example, $\nabla_{\boldsymbol{\theta}} h(\mathbf{X},\boldsymbol{\eta}, \boldsymbol{\theta}^*)$ denotes the gradient of $h(\mathbf{X},\boldsymbol{\eta}, \boldsymbol{\theta})$ with respect to $\boldsymbol{\theta}$ at $\boldsymbol{\theta}^*$.


{\bf Assumptions}
\begin{itemize}
	\item[(A)] For $\boldsymbol{\theta}^* := \underset{\boldsymbol{\theta}\in \boldsymbol{\Theta}_d}{\arg\!\min}\, \mathcal{C}(\boldsymbol{\theta})$,
	\begin{itemize}
		\item[(A1)] the optimal distribution
		$Q^* = Q_{\boldsymbol{\theta}^*}$.
		\item[(A2)] $\boldsymbol{\theta}^*$ is an interior point of $\boldsymbol{\Theta}_d$.
		\item[(A3)] $\boldsymbol{\Gamma} := -\mathbb{E}_{\mu}\!\left[r  \nabla_{\boldsymbol{\theta}}^2 \log{q_{\boldsymbol{\theta}^*}} \right]$ is nonsingular.
		\item[(A4)] $\nabla_{\boldsymbol{\theta}}\!\left( 1 / q_{\boldsymbol{\theta}}(\mathbf{x})\right)$ is continuous in a neighborhood (under the Euclidean norm) of $\boldsymbol{\theta}^*$ for a.e. $\mathbf{x}$ under $\mu$.
		\item[(A5)] $\left\Vert  \mathbb{E}_{\mu}\!\left[ \nabla_{\boldsymbol{\theta}}\!\left( 1 / q_{\boldsymbol{\theta}}\right) \nabla_{\boldsymbol{\theta}}q_{\boldsymbol{\theta}^*} \left(\nabla_{\boldsymbol{\theta}}q_{\boldsymbol{\theta}^*}\right)^T \right]  \right\Vert$ is bounded 
		for all $\boldsymbol{\theta}\in \boldsymbol{\Theta}_d$
		and $\boldsymbol{\Theta}_d$ is a compact set.
		\item[(A6)] There exists a measurable function $g_0(\mathbf{x})$ such that  $\int g_0  \,\mathrm{d}\mu < \infty$ and 
		$$
		\max\{\left\Vert \nabla_{\boldsymbol{\theta}}  q_{\boldsymbol{\theta}}(\mathbf{x})\right\Vert,
		\left\Vert \nabla^2_{\boldsymbol{\theta}}  q_{\boldsymbol{\theta}}(\mathbf{x})\right\Vert \} \le  g_0(\mathbf{x})$$
		for all $\boldsymbol{\theta}\in \boldsymbol{\Theta}_d$.
	\end{itemize}
	\item[(B)] For any $\boldsymbol{\eta} \in \boldsymbol{\Theta}_d$, 
	\begin{itemize}
		\item[(B1)] $h(\mathbf{x}, \boldsymbol{\eta}, \cdot)$ is continuous on $\boldsymbol{\Theta}_d$ for a.e. $\mathbf{x}$ under $Q_{\boldsymbol{\eta}}$.
		\item[(B2)] there exists a measurable function $g_{\boldsymbol{\eta}}$ such that $\int |g_{\boldsymbol{\eta}}| \,\mathrm{d}Q_{\boldsymbol{\eta}} < \infty$ and
		$\left\vert h(\mathbf{x}, \boldsymbol{\eta}, \boldsymbol{\theta}) \right\vert \le g_{\boldsymbol{\eta}}(\mathbf{x})$ for  a.e. $\mathbf{x}$ under $Q_{\boldsymbol{\eta}}$ and all $\boldsymbol{\theta} \in \boldsymbol{\Theta}_d$.
		\item[(B3)] $h(\mathbf{x}, \boldsymbol{\eta}, \cdot)$ is thrice continuously differentiable on $\boldsymbol{\Theta}_d$  for a.e. $\mathbf{x}$ under $Q_{\boldsymbol{\eta}}$.
		\item[(B4)] there exist measurable functions $g_{\boldsymbol{\eta}}^{(i)}$ such that $\int |g_{\boldsymbol{\eta}}^{(i)}|^4 \,\mathrm{d}Q_{\boldsymbol{\eta}} < \infty$ and 
		$\left\Vert \nabla_{\boldsymbol{\theta}}^i h(\mathbf{x}, \boldsymbol{\eta}, \boldsymbol{\theta}) \right\Vert_{\max} \le g_{\boldsymbol{\eta}}^{(i)}(\mathbf{x})$ for $i =1,2,3$ 
		and all $\boldsymbol{\theta}  \in \boldsymbol{\Theta}_d$. The norm is
		the elementwise maximal norm. 
		\item[(B5)] $\boldsymbol{\Lambda}_{\boldsymbol{\eta}} := \mathbb{E}_{Q_{\boldsymbol{\eta}}}\!\left[ \nabla_{\boldsymbol{\theta}} h(\mathbf{X},\boldsymbol{\eta}, \boldsymbol{\theta}^*)  \nabla_{\boldsymbol{\theta}} h(\mathbf{X},\boldsymbol{\eta}, \boldsymbol{\theta}^*)^T\right]$ exists
		and $\int\!\left\Vert \nabla_{\boldsymbol{\theta}}^2 h\!\left(\mathbf{x},\boldsymbol{\eta}, \boldsymbol{\theta}^*\right) \right\Vert_{\max}^2 \,\mathrm{d}Q_{\boldsymbol{\eta}}(\mathbf{x})<\infty$.
	\end{itemize}
	
\end{itemize}




Assumptions (A1-3) and (B1-4)
are the standard regularity conditions to establish consistency (see Lemma~\ref{lemma:consistency}) and asymptotic normality (see Lemma~\ref{lemma:asymp_norm}) of an M-estimator. Assumptions (B3) and (B4) imply Assumptions (B1) and (B2), respectively, as Lemma~\ref{lemma:asymp_norm} requires stronger assumptions than Lemma~\ref{lemma:consistency}.  

To establish Lemma~\ref{lem:expectation} and Theorem~\ref{thm:asymp_bias},
additional assumptions (A4-6) and (B5) are needed
because we need to control the expectation of the smaller order 
terms in the asymptotic analysis. 
Assumptions (A4-6)
ensure that we can exchange the expectation and derivative 
of the function $h\!\left(\mathbf{X}, \boldsymbol{\eta}, \boldsymbol{\theta}\right)$. 
Assumption (B5) bounds the second moment 
of the Hessian matrix of $h$.
This will be needed when deriving the convergence of
sample Hessian matrix to the population Hessian matrix.

A crucial difference from the conventional analysis
on the asymptotic normality
is that to derive the AIC and CIC, we need expectation bound on the 
smaller order terms. 
The asymptotic normality only requires convergence in probability and distribution, which are not enough for the AIC and CIC. 
Also, the bounded derivatives up to the third (not second) order in (B4) 
is an additional requirement that allows us to bound the Taylor's remainder term. 
This is needed because there is no multivariate mean value theorem.

Note that assumptions (B4-5) imply the uniform integrability conditions that are commonly used in
the literature for information criteria akin to the AIC to make the model complexity penalty to be expressed in the free parameter dimension $d$ (see Conditions A7--A8 in \citet{donohue2011conditional}, Theorem~1 in \citet{Claeskens2008}, and more references cited in \citet[][p.1157]{bhansali1991convergence} and \citet[][p.416]{findley2002aic}).
Essentially, the uniform integrability condition
is to make sure the expectation of the smaller order terms is
still a smaller order term. 
Here we use bounds on derivatives and Taylor remainder theorem to handle this.

Below, we provide the proofs of Lemma~\ref{lem:expectation} and Theorem~\ref{thm:asymp_bias}.

\begin{proof}[Proof of Lemma~\ref{lem:expectation}]

	We define the directional derivative with respect to variable ${\boldsymbol{\theta}}\in\boldsymbol{\Theta}_d\subset\R^d$ in the direction $v \in \R^d$ as 
	$$
	\nabla_{{\boldsymbol{\theta}},v} := v^T\nabla_{\boldsymbol{\theta}}. 
	$$

	Recall that $\boldsymbol{\theta}^* := \underset{\boldsymbol{\theta}\in \boldsymbol{\Theta}_d}{\arg\!\min}\, \mathcal{C}(\boldsymbol{\theta})$ is the population minimizer and 
	\begin{align*}
	\hat{\boldsymbol{\theta}}_n &=
	\underset{\boldsymbol{\theta}\in \boldsymbol{\Theta}_d}{\arg\!\min}\,\bar{\mathcal{C}}_{\boldsymbol{\eta}}\!\left(\boldsymbol{\theta}\right)\\
	&= \underset{\boldsymbol{\theta}\in \boldsymbol{\Theta}_d}{\arg\!\min}\, \frac{1}{n}\sum_{i=1}^n h(\mathbf{X}_i; {\boldsymbol{\eta}}, {\boldsymbol{\theta}})\\
	& = \underset{\boldsymbol{\theta}\in \boldsymbol{\Theta}_d}{\arg\!\min}\, h_n({\boldsymbol{\eta}},{\boldsymbol{\theta}}) 
	\end{align*}
	is the estimator, where $\mathbf{X}_i\sim Q_{\boldsymbol{\eta}}$ and
	$$
	h_n({\boldsymbol{\eta}},{\boldsymbol{\theta}}) := \frac{1}{n}\sum_{i=1}^n h(\mathbf{X}_i; {\boldsymbol{\eta}}, {\boldsymbol{\theta}}).
	$$
	
	The expectation of $h_n$ is
	$$
	\E_{Q_{\boldsymbol{\eta}}}(h(\mathbf{X}_i; {\boldsymbol{\eta}}, {\boldsymbol{\theta}})) =  \mathcal{C}({\boldsymbol{\theta}})
	$$
	for all $\boldsymbol{\eta}\in \boldsymbol{\Theta}_d.$
	
	Because ${\boldsymbol{\theta}}^*$ and $\hat{\boldsymbol{\theta}}_n$ are the minimizers,
	they solve the score equations 
	$$
	0 = \nabla_{\boldsymbol{\theta}} \mathcal{C}({\boldsymbol{\theta}}^*) = \nabla_{\boldsymbol{\theta}} h_n({\boldsymbol{\eta}},\hat{\boldsymbol{\theta}}_n). 
	$$
	Thus, by the Taylor's remainder theorem, 
	\begin{align*}
	-\nabla_{\boldsymbol{\theta}} h_n({\boldsymbol{\eta}},{\boldsymbol{\theta}}^*) & = \nabla_{\boldsymbol{\theta}} h_n({\boldsymbol{\eta}},\hat{\boldsymbol{\theta}}_n)-\nabla_{\boldsymbol{\theta}} h_n({\boldsymbol{\eta}},{\boldsymbol{\theta}}^*)\\
	& = \nabla_{{\boldsymbol{\theta}}, \hat{\boldsymbol{\theta}}_n-{\boldsymbol{\theta}}^*} \nabla_{\boldsymbol{\theta}} h_n({\boldsymbol{\eta}},{\boldsymbol{\theta}}^*) + \int_{\epsilon=0}^{\epsilon=1} \nabla_{{\boldsymbol{\theta}}, \hat{\boldsymbol{\theta}}_n-{\boldsymbol{\theta}}^*} \nabla_{{\boldsymbol{\theta}}, \hat{\boldsymbol{\theta}}_n-{\boldsymbol{\theta}}^*} \nabla_{\boldsymbol{\theta}} h_n({\boldsymbol{\eta}},{\boldsymbol{\theta}}^*+\epsilon(\hat{\boldsymbol{\theta}}_n-{\boldsymbol{\theta}}^*))\,\mathrm{d}\epsilon\\
	& = \boldsymbol{\Gamma}_n({\boldsymbol{\eta}},{\boldsymbol{\theta}}^*)(\hat{\boldsymbol{\theta}}_n-{\boldsymbol{\theta}}^*) +R_{1,n}(\boldsymbol{\eta}),
	\end{align*}
	where 
	\begin{align*}
	\boldsymbol{\Gamma}_n({\boldsymbol{\eta}},{\boldsymbol{\theta}}^*)&:=\nabla_{\boldsymbol{\theta}}^2 h_n({\boldsymbol{\eta}},{\boldsymbol{\theta}}^*),\\
	R_{1,n}(\boldsymbol{\eta}) &:= \int_{\epsilon=0}^{\epsilon=1} \nabla_{{\boldsymbol{\theta}}, \hat{\boldsymbol{\theta}}_n-{\boldsymbol{\theta}}^*} \nabla_{{\boldsymbol{\theta}}, \hat{\boldsymbol{\theta}}_n-{\boldsymbol{\theta}}^*} \nabla_{\boldsymbol{\theta}} h_n({\boldsymbol{\eta}},{\boldsymbol{\theta}}^*+\epsilon(\hat{\boldsymbol{\theta}}_n-{\boldsymbol{\theta}}^*))\,\mathrm{d}\epsilon.
	\end{align*}
	$R_{1,n}(\boldsymbol{\eta})$ represents the remainder terms. 
	
	Thus, when $\boldsymbol{\Gamma}_n({\boldsymbol{\eta}},{\boldsymbol{\theta}}^*)$ is invertible, the difference in the estimator can be expressed as 
	\begin{equation}
	\sqrt{n}(\hat{\boldsymbol{\theta}}_n-{\boldsymbol{\theta}}^*) = -\sqrt{n}\boldsymbol{\Gamma}^{-1}_n({\boldsymbol{\eta}},{\boldsymbol{\theta}}^*) \nabla_{\boldsymbol{\theta}} h_n({\boldsymbol{\eta}},{\boldsymbol{\theta}}^*)-\sqrt{n}\boldsymbol{\Gamma}^{-1}_n({\boldsymbol{\eta}},{\boldsymbol{\theta}}^*)R_{1,n}(\boldsymbol{\eta}).
	\label{eq::asymp}
	\end{equation}
	The first quantity is how we establish the asymptotic normality
	and the second part (remainder term) is something we need to bound its expectation.
	
	{\bf Bounding the remainder term $R_{1,n}(\boldsymbol{\eta})$.}
	To see how we obtain this, first note that 
	\begin{equation}
	\E(\|\boldsymbol{\Gamma}^{-1}_n({\boldsymbol{\eta}},{\boldsymbol{\theta}}^*)R_{1,n}(\boldsymbol{\eta})\|) \leq \sqrt{\E(\|\boldsymbol{\Gamma}^{-1}_n({\boldsymbol{\eta}},{\boldsymbol{\theta}}^*)\|^2)\E(\|R_{1,n}(\boldsymbol{\eta})\|^2)}.
	\label{eq::bound1}
	\end{equation}
	Here both norms are the $2$-norm for matrices and vectors.
	So it suffices to bound the two terms individually.
	
	
	{\bf Bounding $\E(\|\boldsymbol{\Gamma}^{-1}_n({\boldsymbol{\eta}},{\boldsymbol{\theta}}^*)\|^2)$. }
	To bound the inverse matrix part, 
	recall that $\boldsymbol{\Gamma}= \nabla_{\boldsymbol{\theta}}^2 \mathcal{C}(\boldsymbol{\theta}^*)$ and note that since $\E(\boldsymbol{\Gamma}_n({\boldsymbol{\eta}},{\boldsymbol{\theta}}^*)) = \boldsymbol{\Gamma}$ due to assumption (A6)
	and $\boldsymbol{\Gamma}_n({\boldsymbol{\eta}},{\boldsymbol{\theta}}^*)$ is essentially a sample mean matrix so 
	the variance shrinks at rate $O(1/n)$ due to assumption (B5). 
	Thus, we can decompose 
	\begin{equation}
	\boldsymbol{\Gamma}_n({\boldsymbol{\eta}},{\boldsymbol{\theta}}^*) = \boldsymbol{\Gamma}+ \Delta_{n,\boldsymbol{\Gamma}}({\boldsymbol{\eta}})
	\label{eq::gamma}
	\end{equation}
	with $\E(\|\Delta_{n,\boldsymbol{\Gamma}}({\boldsymbol{\eta}})\|) = O(1/\sqrt{n})$.
	Moreover, because assumption (B5) holds for all $\boldsymbol{\eta}\in \boldsymbol{\Theta}_d$, 
	$$
	\sup_{\boldsymbol{\eta}\in \boldsymbol{\Theta}_d}\E(\|\Delta_{n,\boldsymbol{\Gamma}}({\boldsymbol{\eta}})\|)  = O(1/\sqrt{n}).
	$$

	For the inverse matrix, because
	\begin{align*}
	\boldsymbol{\Gamma}^{-1}_n({\boldsymbol{\eta}},{\boldsymbol{\theta}}^*) & = [\boldsymbol{\Gamma}+\Delta_{n,\boldsymbol{\Gamma}}({\boldsymbol{\eta}})]^{-1}\\
	& = [\boldsymbol{\Gamma}(I+\boldsymbol{\Gamma}^{-1}\Delta_{n,\boldsymbol{\Gamma}}({\boldsymbol{\eta}}))]^{-1}\\
	& = (I-\boldsymbol{\Gamma}^{-1}\Delta_{n,\boldsymbol{\Gamma}}({\boldsymbol{\eta}}) + O(\|\Delta_{n,\boldsymbol{\Gamma}}({\boldsymbol{\eta}})\|^2))\boldsymbol{\Gamma}^{-1}\\
	& = \boldsymbol{\Gamma}^{-1}-\boldsymbol{\Gamma}^{-1}\Delta_{n,\boldsymbol{\Gamma}}({\boldsymbol{\eta}})\boldsymbol{\Gamma}^{-1}+ O(\|\Delta_{n,\boldsymbol{\Gamma}}({\boldsymbol{\eta}})\|^2).
	\end{align*}
	By the fact that $\boldsymbol{\Gamma}$ is invertible and $\sup_{\boldsymbol{\eta}\in \boldsymbol{\Theta}_d} \E(\|\Delta_{n,\boldsymbol{\Gamma}}({\boldsymbol{\eta}})\|) = O(1/\sqrt{n})$, 
	we conclude that 
	\begin{equation}
	\sup_{\boldsymbol{\eta}\in \boldsymbol{\Theta}_d} \E(\|\boldsymbol{\Gamma}^{-1}_n({\boldsymbol{\eta}},{\boldsymbol{\theta}}^*)\|) = \|\boldsymbol{\Gamma}^{-1}\| + O(1/\sqrt{n})
	\label{eq::bound::inverse}
	\end{equation}
	and
	\begin{equation}
	\sup_{\boldsymbol{\eta}\in \boldsymbol{\Theta}_d} \E(\|\boldsymbol{\Gamma}^{-1}_n({\boldsymbol{\eta}},{\boldsymbol{\theta}}^*)\|^2) = \|\boldsymbol{\Gamma}^{-1}\|^2 + O(1/\sqrt{n}).
	\label{eq::bound::inverse2}
	\end{equation}
	{\bf Bounding $\E(\|R_{1,n}(\boldsymbol{\eta})\|^2)$. }
	By the definition of $R_{1,n}(\boldsymbol{\eta})$,
	\begin{align*}
	R_{1,n}(\boldsymbol{\eta}) &= \int_{\epsilon=0}^{\epsilon=1} \nabla_{{\boldsymbol{\theta}}, \hat{\boldsymbol{\theta}}_n-{\boldsymbol{\theta}}^*} \nabla_{{\boldsymbol{\theta}}, \hat{\boldsymbol{\theta}}_n-{\boldsymbol{\theta}}^*} \nabla_{\boldsymbol{\theta}} h_n({\boldsymbol{\eta}},{\boldsymbol{\theta}}^*+\epsilon(\hat{\boldsymbol{\theta}}_n-{\boldsymbol{\theta}}^*))\,\mathrm{d}\epsilon
	\end{align*}
	so its norm can be bounded by 
	\begin{align*}
	\|R_{1,n}(\boldsymbol{\eta})\|&\leq \|\hat{\boldsymbol{\theta}}_n-{\boldsymbol{\theta}}^*\|^2 \sup_{{\boldsymbol{\theta}}\in \boldsymbol{\Theta}_d }\sup_{\|v\|=\|u\|=\|w\|=1} \|\nabla_{{\boldsymbol{\theta}},v} \nabla_{{\boldsymbol{\theta}},u} \nabla_{{\boldsymbol{\theta}},w}h_n({\boldsymbol{\eta}},{\boldsymbol{\theta}})\|\\
	&\leq  \|\hat{\boldsymbol{\theta}}_n-{\boldsymbol{\theta}}^*\|^2 d^{3/2} \int g_{\boldsymbol{\eta}}^{(3)} \,\mathrm{d}\hat Q_{\boldsymbol{\eta}}
	\end{align*}
	using assumption (B4) and $d$ is the dimension of ${\boldsymbol{\theta}}$ and $\hat Q_{\boldsymbol{\eta}}$ is the empirical measure formed by $\mathbf{X}_1,\cdots, \mathbf{X}_n$.

	Because assumption (B4) holds for all ${\boldsymbol{\theta}}$ and ${\boldsymbol{\eta}}$, 
	\begin{equation}
	\begin{aligned}
	\sup_{\boldsymbol{\eta}\in \boldsymbol{\Theta}_d} \E\|R_{1,n}(\boldsymbol{\eta})\|^2&\leq  \sup_{\boldsymbol{\eta}\in \boldsymbol{\Theta}_d}d^3\E\left(\|\hat{\boldsymbol{\theta}}_n-{\boldsymbol{\theta}}^*\|^4  \left(\int g_{\boldsymbol{\eta}}^{(3)} \,\mathrm{d}\hat Q_{\boldsymbol{\eta}}\right)^2\right)\\&\leq
	d^3\sqrt{\E\left(\|\hat{\boldsymbol{\theta}}_n-{\boldsymbol{\theta}}^*\|^8\right) \sup_{\boldsymbol{\eta}\in \boldsymbol{\Theta}_d}\E\left( \left(\int g_{\boldsymbol{\eta}}^{(3)} \,\mathrm{d}\hat Q_{\boldsymbol{\eta}}\right)^4\right)}\\
	&= O\left(\sqrt{\E\|\hat{\boldsymbol{\theta}}_n-{\boldsymbol{\theta}}^*\|^8}\right) = O(1/n^2).
	\end{aligned}
	\label{eq::bound::R1}
	\end{equation}
	
	
	Putting it altogether, we obtain 
	\begin{align*}
	\sup_{\boldsymbol{\eta}\in \boldsymbol{\Theta}_d} \sqrt{n}\E(\|\boldsymbol{\Gamma}^{-1}_n({\boldsymbol{\eta}},{\boldsymbol{\theta}}^*)R_{1,n}(\boldsymbol{\eta})\|) &\leq \sup_{\boldsymbol{\eta}\in \boldsymbol{\Theta}_d}\sqrt{n}\sqrt{\E(\|\boldsymbol{\Gamma}^{-1}_n({\boldsymbol{\eta}},{\boldsymbol{\theta}}^*)\|^2)\E(\|R_{1,n}(\boldsymbol{\eta})\|^2)}\\
	& = \sqrt{n}(\|\boldsymbol{\Gamma}^{-1}\| + O(1/n^{1/4})) \cdot O\left(\sqrt{\E\|\hat{\boldsymbol{\theta}}_n-{\boldsymbol{\theta}}^*\|^4}\right)\\
	& = O(1/\sqrt{n}) = o(1).
	\end{align*}
	
	{\bf Analysis of the first order term in equation \eqref{eq::asymp}.}
	Note that the difference between the first order term and $Z_{\boldsymbol{\eta}}$
	is
	$$
	R_{2,n}(\boldsymbol{\eta}) := \sqrt{n}\boldsymbol{\Gamma}^{-1}_n({\boldsymbol{\eta}},{\boldsymbol{\theta}}^*)  \nabla_{\boldsymbol{\theta}}h_n({\boldsymbol{\eta}},{\boldsymbol{\theta}}^*)-Z_{\boldsymbol{\eta}}
	= \sqrt{n}(\boldsymbol{\Gamma}^{-1}_n({\boldsymbol{\eta}},{\boldsymbol{\theta}}^*)-\boldsymbol{\Gamma}^{-1})  \nabla_{\boldsymbol{\theta}}h_n({\boldsymbol{\eta}},{\boldsymbol{\theta}}^*)
	$$
	so we need to show that 
	$$
	\sup_{\boldsymbol{\eta}\in \boldsymbol{\Theta}_d}\E\|R_{2,n}(\boldsymbol{\eta})\| = o(1).
	$$
	We can decompose the above bound via
	\begin{align*}
	\sup_{\boldsymbol{\eta}\in \boldsymbol{\Theta}_d}\E\|R_{2,n}(\boldsymbol{\eta})\|
	&\leq \sup_{\boldsymbol{\eta}\in \boldsymbol{\Theta}_d}\sqrt{\E\left\Vert\boldsymbol{\Gamma}^{-1}_n({\boldsymbol{\eta}},{\boldsymbol{\theta}}^*)-\boldsymbol{\Gamma}^{-1}\right\Vert^2 \cdot n\E\left\Vert\nabla_{\boldsymbol{\theta}}h_n({\boldsymbol{\eta}},{\boldsymbol{\theta}}^*)\right\Vert^2}.
	\end{align*}
	Using the derivation of equation \eqref{eq::bound::inverse},
	we can easily show that 
	$$
	\sup_{\boldsymbol{\eta}\in \boldsymbol{\Theta}_d}\E\|\boldsymbol{\Gamma}^{-1}_n({\boldsymbol{\eta}},{\boldsymbol{\theta}}^*)-\boldsymbol{\Gamma}^{-1}\|^2 = O(1/\sqrt{n})
	$$
	and the second part 
	$$
	n\E\|\nabla_{\boldsymbol{\theta}}h_n({\boldsymbol{\eta}},{\boldsymbol{\theta}}^*)\|^2 = O(1)
	$$
	because the covariance matrix ${\sf Cov}(\nabla_{\boldsymbol{\theta}}h_n({\boldsymbol{\eta}},{\boldsymbol{\theta}}^*)) = \frac{1}{n}\boldsymbol{\Lambda}_{\boldsymbol{\eta}}$,
	where 
	$$
	\boldsymbol{\Lambda}_{\boldsymbol{\eta}} = \mathbb{E}_{Q_{\boldsymbol{\eta}}}\!\left[\nabla_{\boldsymbol{\theta}} h\left(\mathbf{X},\boldsymbol{\eta}, \boldsymbol{\theta}^* \right)  \nabla_{\boldsymbol{\theta}} h\left(\mathbf{X},\boldsymbol{\eta}, \boldsymbol{\theta}^*\right)^T \right]  
	$$
	is a bounded matrix for all $\boldsymbol{\eta}$ due to assumption (B5) and $\E\|\nabla_{\boldsymbol{\theta}}h_n({\boldsymbol{\eta}},{\boldsymbol{\theta}}^*)\|^2 = {\sf Tr}({\sf Cov}(\nabla_{\boldsymbol{\theta}}h_n({\boldsymbol{\eta}},{\boldsymbol{\theta}}^*)))$. ${\sf Tr}(\boldsymbol{A})$ is the trace of the matrix $\boldsymbol{A}$.
	
	Equation \eqref{eq::asymp} also implies a useful result about the limiting behavior of $\sqrt{n}(\hat{\boldsymbol{\theta}}_n-{\boldsymbol{\theta}}^*)$. 
	First, because $\boldsymbol{\Gamma}^{-1}_n({\boldsymbol{\eta}},{\boldsymbol{\theta}}^*) $ converges in probability to $\boldsymbol{\Gamma}^{-1}$
	and $\nabla_{\boldsymbol{\theta}} h_n({\boldsymbol{\eta}},{\boldsymbol{\theta}}^*)$ is a sample average quantity,
	we have a central limit theory about $\sqrt{n}(\hat{\boldsymbol{\theta}}_n-{\boldsymbol{\theta}}^*)$ in the sense that
	$$
	\sqrt{n}(\hat{\boldsymbol{\theta}}_n-{\boldsymbol{\theta}}^*) \overset{D}{\rightarrow} N(0, {\boldsymbol{\Sigma}_{\boldsymbol{\eta}}}),
	$$
	where ${\boldsymbol{\Sigma}_{\boldsymbol{\eta}}} := \boldsymbol{\Gamma}^{-1}\boldsymbol{\Lambda}_{\boldsymbol{\eta}} \boldsymbol{\Gamma}^{-1}$
	and ${\boldsymbol{\Sigma}_{\boldsymbol{\eta}}}$ is the asymptotic variance of $\hat{\boldsymbol{\theta}}_n-{\boldsymbol{\theta}}^*$, i.e., 
	$$
	n {\sf Var}(\hat{\boldsymbol{\theta}}_n-{\boldsymbol{\theta}}^*) \rightarrow {\boldsymbol{\Sigma}_{\boldsymbol{\eta}}}.
	$$
	Putting it altogether,
	we conclude that 
	$$
	\sup_{\boldsymbol{\eta}\in \boldsymbol{\Theta}_d}\E\|R_{2,n}(\boldsymbol{\eta})\| = O(1/n^{1/4})
	$$
	and we can rewrite equation \eqref{eq::asymp}
	as 
	\begin{align*}
	\sqrt{n}(\hat{\boldsymbol{\theta}}_n-{\boldsymbol{\theta}}^*) &= -\sqrt{n}\boldsymbol{\Gamma}^{-1}_n({\boldsymbol{\eta}},{\boldsymbol{\theta}}^*) \nabla_{\boldsymbol{\theta}} h_n({\boldsymbol{\eta}},{\boldsymbol{\theta}}^*)-\sqrt{n}\boldsymbol{\Gamma}^{-1}_n({\boldsymbol{\eta}},{\boldsymbol{\theta}}^*)R_{1,n}(\boldsymbol{\eta})\\
	& = -Z_{\boldsymbol{\eta}}  \underbrace{-\sqrt{n}\boldsymbol{\Gamma}^{-1}_n({\boldsymbol{\eta}},{\boldsymbol{\theta}}^*) R_{1,n}(\boldsymbol{\eta}) - R_{2,n}(\boldsymbol{\eta})}_{:=\epsilon_{n,\boldsymbol{\eta}}}
	\end{align*}
	with $\sup_{\boldsymbol{\eta}\in \boldsymbol{\Theta}_d}\E(\|\epsilon_{n,\boldsymbol{\eta}}\|) = O(1/n^{1/4})$.
	
	The covariance of $Z_{\boldsymbol{\eta}}$
	is 
	$$
	{\sf Cov}(Z_{\boldsymbol{\eta}}) =\boldsymbol{\Gamma}^{-1}\boldsymbol{\Lambda}_{\boldsymbol{\eta}}  \boldsymbol{\Gamma}^{-1} 
	$$
	and we have
	\begin{align}
	\boldsymbol{\Lambda}_{\boldsymbol{\eta}} &= \mathbb{E}_{Q_{\boldsymbol{\eta}}}\!\left[\nabla_{\boldsymbol{\theta}} h\left(\mathbf{X},\boldsymbol{\eta}, \boldsymbol{\theta}^* \right)  \nabla_{\boldsymbol{\theta}} h\left(\mathbf{X},\boldsymbol{\eta}, \boldsymbol{\theta}^*\right)^T \right]   \nonumber
	\\&= \mathbb{E}_{Q_{\boldsymbol{\eta}}}\!\left[ \frac{r^2(\mathbf{X})}{q^2_{\boldsymbol{\eta}}\!\left(\mathbf{X} \right)} \nabla_{\boldsymbol{\theta}}\log{q_{\boldsymbol{\theta}^*}\!\left(\mathbf{X} \right)}  \left(\nabla_{\boldsymbol{\theta}}\log{q_{\boldsymbol{\theta}^*}\!\left(\mathbf{X} \right)}\right)^T \right]   \nonumber
	\\&= \mathbb{E}_{\mu}\!\left[ \frac{r^2}{q_{\boldsymbol{\eta}}} \nabla_{\boldsymbol{\theta}}\log{q_{\boldsymbol{\theta}^*}}  \left(\nabla_{\boldsymbol{\theta}}\log{q_{\boldsymbol{\theta}^*}}\right)^T \right]   \nonumber
	\\&= \mathbb{E}_{\mu}\!\left[ \frac{r^2}{q_{\boldsymbol{\eta}}q^2_{\boldsymbol{\theta}^*}} \nabla_{\boldsymbol{\theta}}q_{\boldsymbol{\theta}^*} \left(\nabla_{\boldsymbol{\theta}}q_{\boldsymbol{\theta}^*}\right)^T \right]   \nonumber
	\\&= \rho^2 \mathbb{E}_{\mu}\!\left[ \frac{1}{q_{\boldsymbol{\eta}}} \nabla_{\boldsymbol{\theta}}q_{\boldsymbol{\theta}^*} \left(\nabla_{\boldsymbol{\theta}}q_{\boldsymbol{\theta}^*}\right)^T \right]   \label{eq:sig_theta_hat_rho_sq}
	\\&= \rho^2 \int \left( \frac{1}{q_{\boldsymbol{\theta}^*}(\mathbf{x})} + R_{3,n}(\mathbf{x}; \boldsymbol{\eta})   \right) \nabla_{\boldsymbol{\theta}}q_{\boldsymbol{\theta}^*}(\mathbf{x}) \left(\nabla_{\boldsymbol{\theta}}q_{\boldsymbol{\theta}^*}(\mathbf{x})\right)^T  \,\mathrm{d}\mu(\mathbf{x})     \label{eq:taylor_q_q^*}
	\\&= \rho \boldsymbol{\Gamma} + \rho^2 \int  R_{3,n}(\mathbf{x}; \boldsymbol{\eta})  
	\nabla_{\boldsymbol{\theta}}q_{\boldsymbol{\theta}^*}(\mathbf{x}) \left(\nabla_{\boldsymbol{\theta}}q_{\boldsymbol{\theta}^*}(\mathbf{x})\right)^T  \,\mathrm{d}\mu(\mathbf{x})  \nonumber 
	\\&\overset{(A5)}{=} \rho \boldsymbol{\Gamma} +  O(\|\boldsymbol{\eta} - \boldsymbol{\theta}^*\|) \label{eq:op1_sig_to_gamma} , 
	\end{align}
	where
	$$
	R_{3,n}(\mathbf{x}; \boldsymbol{\eta})   := \int_{\epsilon=0}^{\epsilon=1}\left(\boldsymbol{\eta} - \boldsymbol{\theta}^* \right)^T \nabla_{\boldsymbol{\theta}}\!\left( \frac{1}{q_{\boldsymbol{\theta}^* + \epsilon(\boldsymbol{\eta}-\boldsymbol{\theta}^*)}(\mathbf{x})} \right)\,\mathrm{d}\epsilon
	$$
	is the Taylor's remainder term.
	
	Therefore,
	$$
	n {\sf Var}(\hat{\boldsymbol{\theta}}_n-{\boldsymbol{\theta}}^*) = \boldsymbol{\Sigma}_{\boldsymbol{\eta}} +o(1) = \rho \boldsymbol{\Gamma}^{-1} +o(1) + O(\|\boldsymbol{\eta} - \boldsymbol{\theta}^*\|).
	$$
	
\end{proof}

\begin{proof}[Proof of Theorem~\ref{thm:asymp_bias}]
	
	We first simplify $\mathcal{C}\!\left(\boldsymbol{\hat{\theta}}_n^{(t)}\right)$ and $\bar{\mathcal{C}}_{\boldsymbol{\hat{\theta}}_n^{(t-1)}}\!\left(\boldsymbol{\hat{\theta}}_n^{(t)}\right)$. We then use the simplified expressions to derive the bias of interest. 
	The first quantity is a population objective function while the second quantity is a sample objective function. 
	
	{\bf Analysis on the population objective function  $\mathcal{C}\!\left(\boldsymbol{\hat{\theta}}_n^{(t)}\right)$.}
	\begin{align}
	\mathcal{C}\!\left(\boldsymbol{\hat{\theta}}_n^{(t)}\right) &= - \int r \log{q_{\boldsymbol{\hat{\theta}}_n^{(t)}}} \,\mathrm{d}\mu   \nonumber
	\\&= - \int \frac{r}{q_{\boldsymbol{\hat{\theta}}_n^{(t-1)}}} q_{\boldsymbol{\hat{\theta}}_n^{(t-1)}} \log{q_{\boldsymbol{\hat{\theta}}_n^{(t)}}} \,\mathrm{d}\mu   \label{eq:q_theta=0=>r=0}
	\\&= - \mathbb{E}_{Q_{\boldsymbol{\hat{\theta}}_n^{(t-1)}}}\!\left[ h\!\left(\mathbf{X},\boldsymbol{\hat{\theta}}_n^{(t-1)}, \boldsymbol{\hat{\theta}}_n^{(t)}\right)\right] ,\nonumber
	\end{align}
	where $\mathbf{X} \sim Q_{\boldsymbol{\hat{\theta}}_n^{(t-1)}}$. The equation in \eqref{eq:q_theta=0=>r=0} holds under the condition $Q^* \ll Q_{\boldsymbol{\theta}}$ for all $\boldsymbol{\theta} \in \boldsymbol{\Theta}_d$, because $q_{\boldsymbol{\hat{\theta}}_n^{(t-1)}}(\mathbf{x}) = 0$ implies $r(\mathbf{x}) = 0$ for any $\mathbf{x}$.
	
	We take a second-order expansion of $\mathcal{C}\!\left(\boldsymbol{\hat{\theta}}_n^{(t)}\right)$ about $\boldsymbol{\theta}^*$
	and use the Taylor's remainder theorem: 
	\begin{align}
	\mathcal{C}\!\left(\boldsymbol{\hat{\theta}}_n^{(t)}\right) &= - \mathbb{E}_{Q_{\boldsymbol{\hat{\theta}}_n^{(t-1)}}} \left[   h\!\left(\mathbf{X},\boldsymbol{\hat{\theta}}_n^{(t-1)}, \boldsymbol{\theta}^*\right)\right]
	- \left(\boldsymbol{\hat{\theta}}_n^{(t)} - \boldsymbol{\theta}^*\right)^T \mathbb{E}_{Q_{\boldsymbol{\hat{\theta}}_n^{(t-1)}}} \left[ \nabla_{\boldsymbol{\theta}} h\!\left(\mathbf{X},\boldsymbol{\hat{\theta}}_n^{(t-1)}, \boldsymbol{\theta}^*\right) \right]  \nonumber
	\\&\;\;\; -   \frac{1}{2} \left(\boldsymbol{\hat{\theta}}_n^{(t)} - \boldsymbol{\theta}^*\right)^T 
	\mathbb{E}_{Q_{\boldsymbol{\hat{\theta}}_n^{(t-1)}}} \left[\nabla_{\boldsymbol{\theta}}^2 h\!\left(\mathbf{X},\boldsymbol{\hat{\theta}}_n^{(t-1)}, \boldsymbol{{\theta}}^*\right)  \right]
	\left(\boldsymbol{\hat{\theta}}_n^{(t)} - \boldsymbol{\theta}^*\right) \nonumber\\  
	&+ R_{4,n}   ,  \label{eq:C_theta_t_2nd_order_exp}
	\end{align}
	where $R_{4,n}$ is the Taylor's remainder term at rate $\|\boldsymbol{\hat{\theta}}_n^{(t)} - \boldsymbol{\theta}^*\|^3$
	and has a bounded expectation of rate $\E(|R_{4,n}|) = O(\E\|\boldsymbol{\hat{\theta}}_n^{(t)} - \boldsymbol{\theta}^*\|^3)$
	due to assumption (B4) on the third derivatives.
	In \eqref{eq:C_theta_t_2nd_order_exp}, the zeroth-order term is $\mathcal{C}\!\left(\boldsymbol{\theta}^*\right)$ by definition and the first-order term is zero because 
	\begin{align}
	\mathbb{E}_{Q_{\boldsymbol{\hat{\theta}}_n^{(t-1)}}} \left[ \nabla_{\boldsymbol{\theta}} h\!\left(\mathbf{X},\boldsymbol{\hat{\theta}}_n^{(t-1)}, \boldsymbol{\theta}^*\right) \right] &= 	\mathbb{E}_{Q_{\boldsymbol{\hat{\theta}}_n^{(t-1)}}} \left[  \frac{r(\mathbf{X})}{q_{\boldsymbol{\hat{\theta}}_n^{(t-1)}}\!\left(\mathbf{X} \right)}   \nabla_{\boldsymbol{\theta}} \log{q_{\boldsymbol{\theta}^*}\!\left(\mathbf{X} \right)} \right]  \nonumber
	\\&=\mathbb{E}_{\mu} \left[  \frac{r(\mathbf{X})}{q_{\boldsymbol{\theta}^*}\!\left(\mathbf{X} \right)}   \nabla_{\boldsymbol{\theta}} q_{\boldsymbol{\theta}^*}\!\left(\mathbf{X} \right) \right]  \nonumber
	\\&=\rho \mathbb{E}_{\mu} \left[     \nabla_{\boldsymbol{\theta}} q_{\boldsymbol{\theta}^*}\!\left(\mathbf{X} \right) \right]  \label{eq:q^*=q_theta^*}
	\\&=\rho  \nabla_{\boldsymbol{\theta}} \mathbb{E}_{\mu} \left[     q_{\boldsymbol{\theta}^*}\!\left(\mathbf{X} \right) \right]  \label{eq:interchange_exp_diff}
	\\&=0 , \nonumber
	\end{align}
	where the equation in \eqref{eq:q^*=q_theta^*} holds under assumption (A1)
	and the interchange of expectation and differentiation in \eqref{eq:interchange_exp_diff} holds under assumption~(A6). 

	Let $\boldsymbol{\delta}_n := \sqrt{n}\left(\boldsymbol{\hat{\theta}}_n^{(t)} -\boldsymbol{\theta}^*\right)$. Using the fact that 
	$\mathbb{E}_{Q_{\boldsymbol{\hat{\theta}}_n^{(t-1)}}} \left[ h\!\left(\mathbf{X},\boldsymbol{\hat{\theta}}_n^{(t-1)}, \boldsymbol{{\theta}}^*\right)  \right] = \mathcal{C}\!\left(\boldsymbol{{\theta}}^*\right)$ implies 
	$$
	\mathbb{E}_{Q_{\boldsymbol{\hat{\theta}}_n^{(t-1)}}} \left[\nabla_{\boldsymbol{\theta}}^2 h\!\left(\mathbf{X},\boldsymbol{\hat{\theta}}_n^{(t-1)}, \boldsymbol{{\theta}}^*\right)  \right]= -\boldsymbol{\Gamma}
	$$
	under assumption (A6),
	we simplify the expression of $\mathcal{C}\!\left(\boldsymbol{\hat{\theta}}_n^{(t)}\right)$ in \eqref{eq:C_theta_t_2nd_order_exp} as
	\begin{align}
	\mathcal{C}\!\left(\boldsymbol{\hat{\theta}}_n^{(t)}\right) &=  \mathcal{C}\!\left(\boldsymbol{\theta}^*\right)
	-   \frac{1}{2n} \boldsymbol{\delta}_n^T 
	\mathbb{E}_{Q_{\boldsymbol{\hat{\theta}}_n^{(t-1)}}} \left[\nabla_{\boldsymbol{\theta}}^2 h\!\left(\mathbf{X},\boldsymbol{\hat{\theta}}_n^{(t-1)}, \boldsymbol{{\theta}}^*\!\right)  \right]
	\boldsymbol{\delta}_n  + R_{4,n}     \nonumber
	\\&=  \mathcal{C}\!\left(\boldsymbol{\theta}^*\right)
	+   \frac{1}{2n} \boldsymbol{\delta}_n^T \boldsymbol{{\Gamma}}
	\boldsymbol{\delta}_n + R_{4,n}  .  \nonumber
	\end{align}
	
	Using Lemma~\ref{lem:expectation}, 
	we can rewrite $\boldsymbol{\delta}_n$ as
	$$
	\boldsymbol{\delta}_n = \sqrt{n}\left(\boldsymbol{\hat{\theta}}^{(t)}_n-\boldsymbol{\theta}^*\right) = Z_{\boldsymbol{\hat{\theta}}^{(t-1)}_n} + \epsilon_{n,\boldsymbol{\hat{\theta}}^{(t-1)}_n},
	$$
	where  $E\|\epsilon_{n,\boldsymbol{\hat{\theta}}^{(t-1)}_n}\| = o(1)$ since we have the uniform bound.
	And the covariance
	\begin{equation}
	\begin{aligned}
	{\sf Cov}(\boldsymbol{\delta}_n)
	& = \E\!\left({\sf Cov}\!\left(\boldsymbol{\delta}_n|\boldsymbol{\hat{\theta}}^{(t-1)}_n\right)\right) + {\sf Cov}\!\left(\mathbb{E}\!\left(\boldsymbol{\delta}_n|\boldsymbol{\hat{\theta}}^{(t-1)}_n\right)\right)\\
	&  =\boldsymbol{\Gamma}^{-1}\E\!\left(\boldsymbol{\Lambda}_{\boldsymbol{\hat{\theta}}^{(t-1)}_n}\right)\boldsymbol{\Gamma}^{-1} + o(1).
	\end{aligned}
	\label{eq::cov_err1}
	\end{equation}
	
	Equation \eqref{eq::cov_err1} is derived as follows. We first bound the quantity ${\sf Cov}\!\left(\mathbb{E}\!\left(\boldsymbol{\delta}_n|\boldsymbol{\hat{\theta}}^{(t-1)}_n\right)\right)$.
	Note that for a random vector $\mathbf{X} = (X_1,X_2)$, its covariance can be bounded by 
	\begin{align*}
	  |{\sf Cov}(X_1,X_2) |&\leq \sqrt{{\sf Var}(X_1) {\sf Var}(X_2)}\\
	   &\leq \max\{{\sf Var}(X_1),{\sf Var}(X_2)\}\\
	   &\leq \max\{\mathbb{E}(X_1^2), \mathbb{E}(X_2^2)\}\\
	   &\leq \mathbb{E}(\max\{X_1^2,X_2^2\})\\
	   &\leq \mathbb{E}(\|\mathbf{X}\|^2).
	\end{align*}
	The same bound can be established for a random vector of length $d$.
	By setting $\mathbf{X} = \mathbb{E}\!\left(\epsilon_{n,\boldsymbol{\eta} = \boldsymbol{\hat{\theta}}^{(t-1)}_n}|\boldsymbol{\hat{\theta}}^{(t-1)}_n\right)$ and using the fact that $\mathbb{E}\!\left(Z_{ \boldsymbol{\hat{\theta}}^{(t-1)}_n}|\boldsymbol{\hat{\theta}}^{(t-1)}_n\right)=0$, 
	each element in the covariance matrix
	 ${\sf Cov}\!\left(\mathbb{E}\!\left(\boldsymbol{\delta}_n|\boldsymbol{\hat{\theta}}^{(t-1)}_n\right)\right)= 
	 {\sf Cov}\!\left(\mathbb{E}\!\left(\epsilon_{n,\boldsymbol{\eta} = \boldsymbol{\hat{\theta}}^{(t-1)}_n}|\boldsymbol{\hat{\theta}}^{(t-1)}_n\right)\right)$
	is less than or equal to
	\begin{align*}
	\mathbb{E}\left\{
	 \left\Vert\mathbb{E}\!\left(\epsilon_{n, \boldsymbol{\hat{\theta}}^{(t-1)}_n}\mid\boldsymbol{\hat{\theta}}^{(t-1)}_n\right)\right\Vert^2
	 \right\}
	 &\leq\mathbb{E}\left\{
	 \mathbb{E}^2\!\left(\left\Vert\epsilon_{n, \boldsymbol{\hat{\theta}}^{(t-1)}_n}\right\Vert\mid\boldsymbol{\hat{\theta}}^{(t-1)}_n\right)
	 \right\}\\
	 &\leq\sup_{\boldsymbol{\hat{\theta}}^{(t-1)}_n\in \boldsymbol{\Theta}_d} \mathbb{E}^2\!\left(\left\Vert\epsilon_{n, \boldsymbol{\hat{\theta}}^{(t-1)}_n}\right\Vert\mid\boldsymbol{\hat{\theta}}^{(t-1)}_n\right)\\
	 & = \left(\sup_{\boldsymbol{\hat{\theta}}^{(t-1)}_n\in \boldsymbol{\Theta}_d} \mathbb{E}\!\left(\left\Vert\epsilon_{n, \boldsymbol{\hat{\theta}}^{(t-1)}_n}\right\Vert\mid\boldsymbol{\hat{\theta}}^{(t-1)}_n\right)\right)^2\\
	 &\leq O(1/\sqrt{n}).
	\end{align*}
	Note that the last inequality is due to Lemma~\ref{lem:expectation}.
	
	For another quantity,
	equation \eqref{eq:op1_sig_to_gamma} implies 
	$$
	\mathbb{E}\left\Vert\boldsymbol{\Lambda}_{\boldsymbol{\hat{\theta}}^{(t-1)}_n}- \rho \boldsymbol{\Gamma}\right\Vert =  O\!\left(\mathbb{E}\left\Vert\boldsymbol{\hat{\theta}}^{(t-1)}_n- \boldsymbol{{\theta}}^*\right\Vert\right) = o(1).
	$$
	So we conclude 
	\begin{equation}
	{\sf Cov}(\boldsymbol{\delta}_n) = \rho \boldsymbol{\Gamma}^{-1} + o(1).
	\end{equation}
	This establishes the bound in equation \eqref{eq::cov_err1}.

	Let
	$$
	\boldsymbol{\tilde\delta}_n := \boldsymbol{{\Gamma}}^{1/2} \boldsymbol{\delta}_n = \sqrt{n}\boldsymbol{{\Gamma}}^{1/2}\left(\boldsymbol{\hat{\theta}}_n^{(t)}-\boldsymbol{\theta}^*\right).
	$$
	Then we have
	\begin{equation}
	\begin{aligned}
	\frac{1}{2n} 	\mathbb{E}\!\left[\boldsymbol{\delta}_n^T \boldsymbol{{\Gamma}}
	\boldsymbol{\delta}_n  \right]
	& =\frac{1}{2n} 	\mathbb{E}\!\left[\boldsymbol{\tilde\delta}_n^T
	\boldsymbol{\tilde\delta}_n  \right]\\
	& = \frac{1}{2n} {\sf Tr} ({\sf Cov}(\boldsymbol{\tilde\delta}_n))+ o\!\left(\frac{1}{n}\right)\\
	& = \frac{1}{2n }{\sf Tr} (\boldsymbol{{\Gamma}}^{1/2}{\sf Cov}(\boldsymbol{\delta}_n)\boldsymbol{{\Gamma}}^{1/2})+ o\!\left(\frac{1}{n}\right)\\
	& = \frac{1}{2n }\rho d + o\!\left(\frac{1}{n}\right).
	\end{aligned}
	\label{eq::rhod}
	\end{equation}

	As a result, we conclude that
	\begin{align}
	\mathbb{E}\!\left[ \mathcal{C}\!\left(\boldsymbol{\hat{\theta}}_n^{(t)}\right) \right] &=	 \mathcal{C}\!\left(\boldsymbol{\theta}^*\right)
	+   \frac{1}{2n} 	\mathbb{E}\!\left[\boldsymbol{\delta}_n^T \boldsymbol{{\Gamma}}
	\boldsymbol{\delta}_n  \right] +\E(R_{4,n}) \nonumber
	\\&= \mathcal{C}\!\left(\boldsymbol{\theta}^*\right)
	+   \frac{\rho d}{2n}  + o\!\left(\frac{1}{n}\right) . \label{eq:exp_C_theta_hat}
	\end{align}

	{\bf Analysis of the sample objective function $\bar{\mathcal{C}}_{\boldsymbol{\hat{\theta}}_n^{(t-1)}}\!\left(\boldsymbol{\hat{\theta}}_n^{(t)}\right)$.}
	
	We take a second-order expansion of  $\bar{\mathcal{C}}_{\boldsymbol{\hat{\theta}}_n^{(t-1)}}\!\left(\boldsymbol{\hat{\theta}}_n^{(t)}\right)$ about $\boldsymbol{\theta}^*$
	with a Taylor remainder theorem on the third-order:
	\begin{align}
	\bar{\mathcal{C}}_{\boldsymbol{\hat{\theta}}_n^{(t-1)}}\!\left(\boldsymbol{\hat{\theta}}_n^{(t)}\right)  &= -\frac{1}{n  }	\sum_{i=1}^{n}	h\!\left(\mathbf{X}_i^{(t-1)}, \boldsymbol{\hat{\theta}}_n^{(t-1)}, \boldsymbol{\hat{\theta}}_n^{(t)}\right)     \nonumber
	\\&=-\frac{1}{n}	\sum_{i=1}^{n}  \left(  h\!\left(\mathbf{X}_i^{(t-1)}, \boldsymbol{\hat{\theta}}_n^{(t-1)}, \boldsymbol{\theta}^*\right)  + \left(\boldsymbol{\hat{\theta}}_n^{(t)} - \boldsymbol{\theta}^*\right)^T \nabla_{\boldsymbol{\theta}} h\!\left(\mathbf{X}_i^{(t-1)},\boldsymbol{\hat{\theta}}_n^{(t-1)}, \boldsymbol{\theta}^*\right) \right.    \nonumber
	\\&\;\;\; + \left. \frac{1}{2} \left(\boldsymbol{\hat{\theta}}_n^{(t)} - \boldsymbol{\theta}^*\right)^T \nabla_{\boldsymbol{\theta}}^2 h\!\left(\mathbf{X}_i^{(t-1)},\boldsymbol{\hat{\theta}}_n^{(t-1)}, \boldsymbol{{\theta}}^*\right) \left(\boldsymbol{\hat{\theta}}_n^{(t)} - \boldsymbol{\theta}^*\right)  \right) \nonumber\\
	& \;\;\; + R_{5,n}   \label{eq:Cbar_theta_hat_2nd_order}
	\end{align}
	where $R_{5,n}$ is the remainder of the expansion that involves $\|\boldsymbol{\hat{\theta}}_n^{(t)} - \boldsymbol{\theta}^*\|^3$
	and the third-order derivative of $h$. 
	By assumption (B4), this quantity has an expectation $\E(|R_{5,n}|) = O(\E\|\boldsymbol{\hat{\theta}}_n^{(t)} - \boldsymbol{\theta}^*\|^3)$.
	The zeroth-order term is $\bar{\mathcal{C}}_{\boldsymbol{\hat{\theta}}_n^{(t-1)}}\!\left(\boldsymbol{\theta}^*\right)$ by definition. To re-express the first-order term, we use the fact
	\begin{align}
	\nabla_{\boldsymbol{\theta}} \bar{\mathcal{C}}_{\boldsymbol{\hat{\theta}}_n^{(t-1)}}\!\left(\boldsymbol{\hat{\theta}}_n^{(t)}\right) &=0 \nonumber
	\end{align}
	or equivalently
	\begin{align}
	0 &= \frac{1}{n} \sum_{i=1}^{n} \nabla_{\boldsymbol{\theta}} h\!\left(\mathbf{X}_i^{(t-1)},\boldsymbol{\hat{\theta}}_n^{(t-1)}, \boldsymbol{\hat{\theta}}_n^{(t)}\right)  \nonumber
	\\&=\frac{1}{n}	\sum_{i=1}^{n} \nabla_{\boldsymbol{\theta}} h\!\left(\mathbf{X}_i^{(t-1)},\boldsymbol{\hat{\theta}}_n^{(t-1)}, \boldsymbol{\theta}^*\right)
	+  \frac{1}{n} \sum_{i=1}^{n} \nabla_{\boldsymbol{\theta}}^2 h\!\left(\mathbf{X}_i^{(t-1)},\boldsymbol{\hat{\theta}}_n^{(t-1)}, \boldsymbol{\theta}^*\right) \left(\boldsymbol{\hat{\theta}}_n^{(t)} - \boldsymbol{\theta}^*\right) \nonumber
	\\
	&\quad +R_{6,n}\label{eq:zero_score_at_theta_hat}
	\end{align}
	where $R_{6,n}$ is another remainder of the expansion that involves $\|\boldsymbol{\hat{\theta}}_n^{(t)} - \boldsymbol{\theta}^*\|^2$
	and the third-order derivative of $h$. 
	By assumption (B4), this quantity has an expectation $\E(|R_{6,n}|) = O(\E\|\boldsymbol{\hat{\theta}}_n^{(t)} - \boldsymbol{\theta}^*\|^2)$.
	
	Rearranging the equation in \eqref{eq:zero_score_at_theta_hat} yields
	\begin{align}
	-\frac{1}{n}	\sum_{i=1}^{n} \nabla_{\boldsymbol{\theta}} h\!\left(\mathbf{X}_i^{(t-1)},\boldsymbol{\hat{\theta}}_n^{(t-1)}, \boldsymbol{\theta}^*\right)
	&=  \frac{1}{n} \sum_{i=1}^{n} \nabla_{\boldsymbol{\theta}}^2 h\!\left(\mathbf{X}_i^{(t-1)},\boldsymbol{\hat{\theta}}_n^{(t-1)}, \boldsymbol{\theta}^*\right) \left(\boldsymbol{\hat{\theta}}_n^{(t)} - \boldsymbol{\theta}^*\right) +R_{6,n}. \nonumber
	\end{align}
	Plugging this to the equation in \eqref{eq:Cbar_theta_hat_2nd_order} results in
	\begin{align}
	\bar{\mathcal{C}}_{\boldsymbol{\hat{\theta}}_n^{(t-1)}}\!\left(\boldsymbol{\hat{\theta}}_n^{(t)}\right)  &=-\frac{1}{n}	\sum_{i=1}^{n}  \left(  h\!\left(\mathbf{X}_i^{(t-1)}, \boldsymbol{\hat{\theta}}_n^{(t-1)}, \boldsymbol{\theta}^*\right)  + \left(\boldsymbol{\hat{\theta}}_n^{(t)} - \boldsymbol{\theta}^*\right)^T \nabla_{\boldsymbol{\theta}} h\!\left(\mathbf{X}_i^{(t-1)},\boldsymbol{\hat{\theta}}_n^{(t-1)}, \boldsymbol{\theta}^*\right) \right.    \nonumber
	\\&\;\;\; + \left. \frac{1}{2} \left(\boldsymbol{\hat{\theta}}_n^{(t)} - \boldsymbol{\theta}^*\right)^T \nabla_{\boldsymbol{\theta}}^2 h\!\left(\mathbf{X}_i^{(t-1)},\boldsymbol{\hat{\theta}}_n^{(t-1)}, \boldsymbol{\theta}^*\right) \left(\boldsymbol{\hat{\theta}}_n^{(t)} - \boldsymbol{\theta}^*\right)  \right)  +R_{5,n}  \nonumber
	\\&=\bar{\mathcal{C}}_{\boldsymbol{\hat{\theta}}_n^{(t-1)}}\!\left(\boldsymbol{\theta}^*\right)   - \frac{1}{n}\boldsymbol{\delta}_n^T  \left( -\frac{1}{n} \sum_{i=1}^{n} \nabla_{\boldsymbol{\theta}}^2 h\!\left(\mathbf{X}_i^{(t-1)},\boldsymbol{\hat{\theta}}_n^{(t-1)}, \boldsymbol{\theta}^*\right) \right) \boldsymbol{\delta}_n     \nonumber
	\\&\;\;\; + \frac{1}{2n}\boldsymbol{\delta}_n^T  \left( -\frac{1}{n} \sum_{i=1}^{n} \nabla_{\boldsymbol{\theta}}^2 h\!\left(\mathbf{X}_i^{(t-1)},\boldsymbol{\hat{\theta}}_n^{(t-1)}, \boldsymbol{\theta}^*\right) \right)   \boldsymbol{\delta}_n +R_{5,n}+\left(\boldsymbol{\hat{\theta}}_n^{(t)} - \boldsymbol{\theta}^*\right)^TR_{6,n}. \nonumber
	\end{align}
	Let  $\boldsymbol{{\Gamma}}_n\!\left(\boldsymbol{\hat{\theta}}_n^{(t-1)},\boldsymbol{\theta}^*\right) :=  -\frac{1}{n} \sum_{i=1}^{n} \nabla_{\boldsymbol{\theta}}^2 h\!\left(\mathbf{X}_i^{(t-1)},\boldsymbol{\hat{\theta}}_n^{(t-1)}, \boldsymbol{\theta}^*\right)$. We can simplify the above equation as:
	\begin{equation}
	\begin{aligned}
	\bar{\mathcal{C}}_{\boldsymbol{\hat{\theta}}_n^{(t-1)}}\!\left(\boldsymbol{\hat{\theta}}_n^{(t)}\right) &=  \bar{\mathcal{C}}_{\boldsymbol{\hat{\theta}}_n^{(t-1)}}\!\left(\boldsymbol{\theta}^*\right)   - \frac{1}{n}\boldsymbol{\delta}_n^T  \boldsymbol{{\Gamma}}_n\!\left(\boldsymbol{\hat{\theta}}_n^{(t-1)},\boldsymbol{\theta}^*\right)  \boldsymbol{\delta}_n  +  \frac{1}{2n}\boldsymbol{\delta}_n^T  \boldsymbol{{\Gamma}}_n\!\left(\boldsymbol{\hat{\theta}}_n^{(t-1)},\boldsymbol{\theta}^*\right)    \boldsymbol{\delta}_n \\&\quad +R_{5,n}+\left(\boldsymbol{\hat{\theta}}_n^{(t)} - \boldsymbol{\theta}^*\right)^TR_{6,n} \\
	&=  \bar{\mathcal{C}}_{\boldsymbol{\hat{\theta}}_n^{(t-1)}}\!\left(\boldsymbol{\theta}^*\right)   -   \frac{1}{2n}\boldsymbol{\delta}_n^T  \boldsymbol{{\Gamma}}_n\left(\boldsymbol{\hat{\theta}}_n^{(t-1)},\boldsymbol{\theta}^*\right)    \boldsymbol{\delta}_n +R_{5,n}+\left(\boldsymbol{\hat{\theta}}_n^{(t)} - \boldsymbol{\theta}^*\right)^TR_{6,n}
	\end{aligned}	
	\label{eq::sample_obj}
	\end{equation}	
	Note that the first term $\mathbb{E}\!\left[ \bar{\mathcal{C}}_{\boldsymbol{\hat{\theta}}_n^{(t-1)}}\!\left(\boldsymbol{\theta}^*\right)  \right] =   \mathcal{C}\!\left(\boldsymbol{\theta}^*\right)$ 
	and
	by equation \eqref{eq::gamma},
	$$
	\boldsymbol{{\Gamma}}_n\!\left(\boldsymbol{\hat{\theta}}_n^{(t-1)},\boldsymbol{\theta}^*\right) = 	\boldsymbol{{\Gamma}} + \Delta_{n,\boldsymbol{{\Gamma}}}\!\left(\boldsymbol{\hat{\theta}}_n^{(t-1)}\right),
	$$
	where the later analysis in the proof of Lemma~\ref{lem:expectation} showed that
	$$
	\E\!\left(\left\Vert\Delta_{n,\boldsymbol{{\Gamma}}}\!\left(\boldsymbol{\hat{\theta}}_n^{(t-1)}\right)\right\Vert\right)
	= O\!\left(\E\!\left(\left\Vert\boldsymbol{\hat{\theta}}_n^{(t-1)}-\boldsymbol{\theta}^*\right\Vert\right)\right) = O(1/\sqrt{n}).
	$$
	Thus, equation \eqref{eq::rhod} yields that $\E\!\left(\boldsymbol{\delta}_n^T  \boldsymbol{{\Gamma}}_n\!\left(\boldsymbol{\hat{\theta}}_n^{(t-1)},\boldsymbol{\theta}^*\right)    \boldsymbol{\delta}_n\right) = \rho d +o(1)$.
	So the expectation of equation \eqref{eq::sample_obj} becomes
	\begin{equation}
	\begin{aligned}
	\mathbb{E}\!\left[\bar{\mathcal{C}}_{\boldsymbol{\hat{\theta}}_n^{(t-1)}}\!\left(\boldsymbol{\hat{\theta}}_n^{(t)}\right) \right] &= \mathcal{C}\!\left(\boldsymbol{\theta}^*\right) -  \frac{\rho d}{2n}   + o\!\left(\frac{1}{n}\right) .
	\end{aligned}
	\label{eq:exp_C_bar_theta_hat}
	\end{equation}

	Therefore, combining \eqref{eq:exp_C_theta_hat} and \eqref{eq:exp_C_bar_theta_hat}, the bias of interest is
	\begin{align}
	\mathbb{E}\!\left[\bar{\mathcal{C}}_{\boldsymbol{\hat{\theta}}_n^{(t-1)}}\!\left(\boldsymbol{\hat{\theta}}_n^{(t)}\right)  - \mathcal{C}\!\left(\boldsymbol{\hat{\theta}}_n^{(t)}\right) \right] &= \mathcal{C}\!\left(\boldsymbol{\theta}^*\right) - \frac{\rho d}{2n}  - \left( \mathcal{C}\!\left(\boldsymbol{\theta}^*\right)
	+   \frac{\rho d}{2n}  \right)   + o\!\left(\frac{1}{n}\right)  \nonumber
	\\&= - \frac{\rho d}{n}  + o\!\left(\frac{1}{n}\right) . \nonumber
	\end{align}

\end{proof}

\section*{Appendix B: EM Algorithm for Minimizing the Cross-Entropy}
This appendix describes our version of the expectation-maximization (EM) algorithm to minimize the cross-entropy from a Gaussian mixture model (GMM) to a target distribution. 

The GMM density is expressed as 
\begin{align}
q\!\left(\mathbf{x} ; \boldsymbol{\theta}\right) &= \sum_{j=1}^{k} \alpha_j \, q_j\!\left(\mathbf{x} ; \boldsymbol{\mu}_j, \boldsymbol{\Sigma}_j\right) , \label{eq:GMM}
\end{align}
where the component weights, $\alpha_j > 0, j= 1,\ldots,k$, satisfy  $\sum_{j=1}^{k} \alpha_j  = 1$. The $j$th Gaussian component density $q_j$ is parametrized by the mean $\boldsymbol{\mu}_j$ and the covariance $\boldsymbol{\Sigma}_j$. Thus, the model parameter $\boldsymbol{\theta}$ denotes $\left(\alpha_1,\ldots,\alpha_k, \boldsymbol{\mu}_1, \ldots, \boldsymbol{\mu}_k, \boldsymbol{\Sigma}_1,\ldots, \boldsymbol{\Sigma}_k \right)$.

To find the MCE $\boldsymbol{\hat{\theta}}^{(t)}$ of $\boldsymbol{\theta}$, we want to minimize $\bar{\mathcal{C}}^{(t-1)}(\boldsymbol{\theta})$ in \eqref{eq:CE_cumulative} and thus set its gradient to zero:
\begin{align}
-\frac{1}{\sum_{s = 0}^{t-1}n_{s}}\sum_{s = 0}^{t-1} 	\sum_{i=1}^{n_s}	\nabla_{\boldsymbol{\theta}} h\!\left(\mathbf{X}_i^{(s)}, \boldsymbol{\hat{\theta}}^{(s)}, \boldsymbol{\theta}\right)  &= 0. \nonumber
\end{align}
This leads to the following updating equations for our version of the EM algorithm:
\begin{align}
\alpha_j &= \frac{\sum_{s = 0}^{t-1}    \sum_{i=1}^{n_s} \frac{r\!\left(\mathbf{X}_i^{(s)}\right)}{q_{\boldsymbol{\hat{\theta}}^{(s)}}\!\left(\mathbf{X}_i^{(s)} \right)}  \gamma_{ij}^{(s)} }{\sum_{s = 0}^{t-1}    \sum_{i=1}^{n_s} \frac{r\!\left(\mathbf{X}_i^{(s)}\right)}{q_{\boldsymbol{\hat{\theta}}^{(s)}}\!\left(\mathbf{X}_i^{(s)} \right)} } , \label{eq:alpha_updating}
\\
\boldsymbol{\mu}_j &= \frac{\sum_{s = 0}^{t-1}    \sum_{i=1}^{n_s} \frac{r\!\left(\mathbf{X}_i^{(s)}\right)}{q_{\boldsymbol{\hat{\theta}}^{(s)}}\!\left(\mathbf{X}_i^{(s)} \right)} \gamma_{ij}^{(s)} \mathbf{X}_i^{(s)} }{\sum_{s = 0}^{t-1}    \sum_{i=1}^{n_s} \frac{r\!\left(\mathbf{X}_i^{(s)}\right)}{q_{\boldsymbol{\hat{\theta}}^{(s)}}\!\left(\mathbf{X}_i^{(s)} \right)} \gamma_{ij}^{(s)} } , \label{eq:mu_updating}
\\
\boldsymbol{\Sigma}_j &= \frac{\sum_{s = 0}^{t-1}    \sum_{i=1}^{n_s} \frac{r\!\left(\mathbf{X}_i^{(s)}\right)}{q_{\boldsymbol{\hat{\theta}}^{(s)}}\!\left(\mathbf{X}_i^{(s)} \right)}   \gamma_{ij}^{(s)} (\mathbf{X}_i^{(s)} - \boldsymbol{\mu}_j)(\mathbf{X}_i^{(s)} - \boldsymbol{\mu}_j)^T}{\sum_{s = 0}^{t-1}    \sum_{i=1}^{n_s} \frac{r\!\left(\mathbf{X}_i^{(s)}\right)}{q_{\boldsymbol{\hat{\theta}}^{(s)}}\!\left(\mathbf{X}_i^{(s)} \right)}  \gamma_{ij}^{(s)} } , \label{eq:sigma_updating}
\end{align}
\vspace{-1em}
where
\vspace{-1.4em}
\begin{align}
\gamma_{ij}^{(s)} &= \frac{ \alpha_j \, q_j\!\left(\mathbf{X}_i^{(s)} ; \boldsymbol{\mu}_j, \boldsymbol{\Sigma}_j\right)}{\sum_{j'=1}^{k} \alpha_{j'} \, q_{j'}\!\left(\mathbf{X}_i^{(s)} ; \boldsymbol{\mu}_{j'}, \boldsymbol{\Sigma}_{j'}\right)} . \label{eq:gamma_ij}
\end{align} 
The right-hand sides of the updating equations in \eqref{eq:alpha_updating}, \eqref{eq:mu_updating}, and \eqref{eq:sigma_updating} involve $$\boldsymbol{\theta} = \left(\alpha_1,\ldots,\alpha_k, \boldsymbol{\mu}_1, \ldots, \boldsymbol{\mu}_k, \boldsymbol{\Sigma}_1,\ldots, \boldsymbol{\Sigma}_k \right)$$ either explicitly or implicitly through $\gamma_{ij}^{(s)}$. Thus, the equations cannot be analytically solved for $\boldsymbol{\theta}$. Instead, starting with an initial guess of $\boldsymbol{\theta}$, our version of the EM algorithm alternates between the expectation step (computing $\gamma_{ij}^{(s)}$) and the maximization step (updating $\boldsymbol{\theta}$) based on the updating equations until convergence is reached. 

Prior studies \citep{botev2013,Wang2015,kurtz2013} using mixture models for the cross-entropy method for importance sampling (recall that this method is a special case of the procedure in Figure~\ref{fig:MCE_iter} when $r$ is proportional to the optimal importance sampling density) do not iterate their updating equations; instead, they solve them only once when new data are gathered.  
This paper uses the aforementioned EM algorithm (i.e., iterating the updating equations until convergence) within the $t^{th}$ iteration to minimize $\bar{\mathcal{C}}^{(t-1)}(\boldsymbol{\theta})$ in \eqref{eq:CE_cumulative}.  

\section*{Appendix C: Implementation Details of the Numerical Example}
This appendix describes the implementation details of the numerical example in Section~\ref{sec:num_ex} to ensure the reproducibility. For the numerical experiment (with 500 repetitions), we randomly determined $\boldsymbol{\hat{\theta}}^{(0)} = \boldsymbol{\eta}$ by drawing $\mu_1, \ldots, \mu_{30}$ from a standard multivariate Gaussian and setting all $\Sigma_1,\ldots, \Sigma_{30}$ as $3I_{p\times p}$. 

For the implementation of the EM algorithm, we used multiple random initial values of $\boldsymbol{\theta}$ and chose the best minimizer of $\bar{\mathcal{C}}^{(t-1)}(\boldsymbol{\theta})$ in \eqref{eq:CE_cumulative} to reduce the impact of initial guess of $\boldsymbol{\theta}$ on the algorithm's performance and avoid getting stuck with a local minimizer \citep{figueiredo2002}. 
In the $t^{th}$ iteration for $t\ge 1$, we randomly selected $\mu_1, \ldots, \mu_k$ from $\{ \mathbf{X}_i^{(s-1)}:  h\!\left(\mathbf{X}_i^{(s-1)}, \boldsymbol{\hat{\theta}}_n^{(s-1)}, \boldsymbol{\theta}\right) > 0, i=1,\ldots,n; s = 1,\ldots, t \}$ without replacement. However, if the set's cardinality was smaller than $k$, we randomly selected any elements in $\{ \mathbf{X}_i^{(s-1)} : h\!\left(\mathbf{X}_i^{(s-1)}, \boldsymbol{\hat{\theta}}_n^{(s-1)}, \boldsymbol{\theta}\right) = 0, i=1,\ldots,n;  s = 1,\ldots, t\}$ for the remaining parameters. We set $\Sigma_1,\ldots, \Sigma_k$ as $(3/p)\, \textrm{trace}\!\left(cov(\bar{\mathbf{X}}) \right)I_{p\times p}$, where $\bar{\mathbf{X}}$ is the data matrix created by augmenting  $\{ \mathbf{X}_i^{(s-1)} : i=1,\ldots,n; s = 1,\ldots, t\}$, and $cov$ is the sample covariance. We used equal component weights for the initialization, $\alpha_j = 1/k, j= 1,\ldots,k$.


When the number of components, $k$, became large enough to cause an overfitting issue within the EM algorithm, we caught it by monitoring the condition numbers of the Gaussian components' covariances \citep{figueiredo2002}. 
We aborted the EM algorithm when the condition number of any covariance exceeded $10^5$. If we needed to abort most of the EM algorithms that started with different initial parameter guesses (we used the threshold of 5 aborted out of 10), it indicated that $k$ is already too large for the given sample size. 

To check the convergence of the EM algorithm, we checked the reduction of $\bar{\mathcal{C}}^{(t-1)}(\boldsymbol{\theta})$ in \eqref{eq:CE_cumulative}. 
We stopped iterating updating equations in the EM algorithm if the reduction of $\bar{\mathcal{C}}^{(t-1)}(\boldsymbol{\theta})$ was less than 1\% or a specified maximum number of EM iterations, 10, is reached. 


We computed the CIC for $k = k_{\min}, k_{\min}+1, \ldots, k_{\max}$ for the grid search of the minimizer $k^{*(t)}$ in the $t^{th}$ iteration. We set $k_{\min}$ as one for $t=1$ and $\max\!\left(1, k^{*(t-1)}-3\right)$ for $t\ge2$. At $k=k_{\min}$, if all random initializations failed to converge, then we reduced $k_{\min}$ by one. In practice, it is generally unnecessary to increase $k$ up to $k_{\max}$, which is upper bounded by a known function of the sample size, because the overfitting is detected within the EM algorithm. To reduce the grid search time, we computed the moving average (with the window size of four) of the CIC and stopped increasing $k$ when the moving average started to increase. 


\end{document}